\documentclass{aastex63}

\usepackage[version=4]{mhchem} % Package for chemical equation typesetting
\usepackage{xspace}
\newcommand{\nthp}{\ce{N2H+}\xspace}
\newcommand{\ntdp}{\ce{N2D+}\xspace}

\received{March 3, 2021}
\revised{June 16, 2021}
\accepted{June 17, 2021}
\submitjournal{ApJ}

\shorttitle{MAPS XIII: Disk Ionization Structure}
\shortauthors{Aikawa et al.}
\graphicspath{{./}{figures/}}
\usepackage[version=4]{mhchem}

\begin{document}

\title{Molecules with ALMA at Planet-forming Scales (MAPS) XIII: HCO$^+$ and disk ionization structure}

\correspondingauthor{Yuri Aikawa}
\email{aikawa@astron.s.u-tokyo.ac.jp}

\author[0000-0003-3283-6884]{Yuri Aikawa}
\affiliation{Department of Astronomy, Graduate School of Science, The University of Tokyo, 113-0033, Tokyo, Japan}

\author[0000-0002-2700-9676]{Gianni Cataldi}
\affiliation{Department of Astronomy, Graduate School of Science, The University of Tokyo, 113-0033, Tokyo, Japan}
\affiliation{National Astronomical Observatory of Japan, 2-21-1 Osawa, Mitaka, Tokyo 181-8588, Japan}

\author[0000-0003-4099-6941]{Yoshihide Yamato}
\affiliation{Department of Astronomy, Graduate School of Science, The University of Tokyo, 113-0033, Tokyo, Japan}

\author[0000-0002-0661-7517]{Ke Zhang}
\altaffiliation{NASA Hubble Fellow}
\affiliation{Department of Astronomy, University of Wisconsin-Madison, 475 N Charter St, Madison, WI 53706}
\affiliation{Department of Astronomy, University of Michigan, 323 West Hall, 1085 S. University Avenue, Ann Arbor, MI 48109, USA}

\author[0000-0003-2014-2121]{Alice S. Booth}
\affiliation{Leiden Observatory, Leiden University, 2300 RA Leiden, the Netherlands}
\affiliation{School of Physics and Astronomy, University of Leeds, Leeds, LS2 9JT, UK}

\author[0000-0002-2026-8157]{Kenji Furuya}
\affiliation{National Astronomical Observatory of Japan, 2-21-1 Osawa, Mitaka, Tokyo 181-8588, Japan}

\author[0000-0003-2253-2270]{Sean M. Andrews}
\affiliation{Center for Astrophysics \textbar\, Harvard \& Smithsonian, 60 Garden St., Cambridge, MA 02138, USA}

\author[0000-0001-7258-770X]{Jaehan Bae}
\altaffiliation{NASA Hubble Fellowship Program Sagan Fellow}
\affil{Earth and Planets Laboratory, Carnegie Institution for Science, 5241 Broad Branch Road NW, Washington, DC 20015, USA}
\affiliation{Department of Astronomy, University of Florida, Gainesville, FL 32611, USA}

\author[0000-0003-4179-6394]{Edwin A.\ Bergin}
\affiliation{Department of Astronomy, University of Michigan, 323 West Hall, 1085 S. University Avenue, Ann Arbor, MI 48109, USA}

\author[0000-0002-8716-0482]{Jennifer B. Bergner}
\altaffiliation{NASA Hubble Fellowship Program Sagan Fellow}
\affiliation{University of Chicago Department of the Geophysical Sciences, Chicago, IL 60637, USA}

\author[0000-0003-4001-3589]{Arthur D. Bosman}
\affiliation{Department of Astronomy, University of Michigan, 323 West Hall, 1085 S. University Avenue, Ann Arbor, MI 48109, USA}

\author[0000-0003-2076-8001]{L. Ilsedore Cleeves}
\affiliation{Department of Astronomy, University of Virginia, Charlottesville, VA 22904, USA}

\author[0000-0002-1483-8811]{Ian Czekala}
\altaffiliation{NASA Hubble Fellowship Program Sagan Fellow}
\affiliation{Department of Astronomy and Astrophysics, 525 Davey Laboratory, The Pennsylvania State University, University Park, PA 16802, USA}
\affiliation{Center for Exoplanets and Habitable Worlds, 525 Davey Laboratory, The Pennsylvania State University, University Park, PA 16802, USA}
\affiliation{Center for Astrostatistics, 525 Davey Laboratory, The Pennsylvania State University, University Park, PA 16802, USA}
\affiliation{Institute for Computational \& Data Sciences, The Pennsylvania State University, University Park, PA 16802, USA}   
\affiliation{Department of Astronomy, 501 Campbell Hall, University of California, Berkeley, CA 94720-3411, USA}

\author[0000-0003-4784-3040]{Viviana V. Guzm\'{a}n}
\affiliation{Instituto de Astrof\'isica, Pontificia Universidad Cat\'olica de Chile, Av. Vicu\~na Mackenna 4860, 7820436 Macul, Santiago, Chile}

\author[0000-0001-6947-6072]{Jane Huang}
\altaffiliation{NASA Hubble Fellowship Program Sagan Fellow}
\affiliation{Department of Astronomy, University of Michigan, 323 West Hall, 1085 S. University Avenue, Ann Arbor, MI 48109, USA}
\affiliation{Center for Astrophysics \textbar\, Harvard \& Smithsonian, 60 Garden St., Cambridge, MA 02138, USA}

\author[0000-0003-1008-1142]{John~D.~Ilee} 
\affiliation{School of Physics and Astronomy, University of Leeds, Leeds, LS2 9JT, UK}

\author[0000-0003-1413-1776]{Charles J. Law}
\affiliation{Center for Astrophysics \textbar\, Harvard \& Smithsonian, 60 Garden St., Cambridge, MA 02138, USA}

\author[0000-0003-1837-3772]{Romane Le Gal}
\affiliation{Center for Astrophysics \textbar\, Harvard \& Smithsonian, 60 Garden St., Cambridge, MA 02138, USA}
\affiliation{IRAP, Universit\'{e} de Toulouse, CNRS, CNES, UT3, 31400 Toulouse, France}
\affiliation{Univ. Grenoble Alpes, CNRS, IPAG, F-38000 Grenoble, France}
\affiliation{IRAM, 300 rue de la piscine, F-38406 Saint-Martin d'H\`{e}res, France}

\author[0000-0002-8932-1219]{Ryan A. Loomis}
\affiliation{National Radio Astronomy Observatory, 520 Edgemont Rd., Charlottesville, VA 22903, USA}

\author[0000-0002-1637-7393]{Fran\c cois M\'enard}\affiliation{Univ. Grenoble Alpes, CNRS, IPAG, F-38000 Grenoble, France}

\author[0000-0002-7058-7682]{Hideko Nomura}
\affiliation{National Astronomical Observatory of Japan, 2-21-1 Osawa, Mitaka, Tokyo 181-8588, Japan}

\author[0000-0001-8798-1347]{Karin I. \"Oberg}
\affiliation{Center for Astrophysics \textbar\, Harvard \& Smithsonian, 60 Garden St., Cambridge, MA 02138, USA}

\author[0000-0001-8642-1786]{Chunhua Qi}
\affiliation{Center for Astrophysics \textbar\, Harvard \& Smithsonian, 60 Garden St., Cambridge, MA 02138, USA}

\author[0000-0002-6429-9457]{Kamber R. Schwarz}
\altaffiliation{NASA Hubble Fellowship Program Sagan Fellow}
\affiliation{Lunar and Planetary Laboratory, University of Arizona, 1629 E. University Blvd, Tucson, AZ 85721, USA}

\author[0000-0003-1534-5186]{Richard Teague}
\affiliation{Center for Astrophysics \textbar\, Harvard \& Smithsonian, 60 Garden St., Cambridge, MA 02138, USA}

\author[0000-0002-6034-2892]{Takashi Tsukagoshi}
\affiliation{National Astronomical Observatory of Japan, 2-21-1 Osawa, Mitaka, Tokyo 181-8588, Japan}

\author[0000-0001-6078-786X]{Catherine Walsh}
\affiliation{School of Physics and Astronomy, University of Leeds, Leeds, LS2 9JT, UK}

\author[0000-0003-1526-7587]{David J. Wilner}
\affiliation{Center for Astrophysics \textbar\, Harvard \& Smithsonian, 60 Garden St., Cambridge, MA 02138, USA}

%\collaboration{1}{(MAPS collaboration)}



\begin{abstract}

We observed HCO$^+$ $J=1-0$ and H$^{13}$CO$^+$ $J=1-0$ emission towards the five protoplanetary disks around IM~Lup, GM~Aur, AS~209, HD~163296, and MWC~480 as part of the MAPS project. HCO$^+$ is detected and mapped at 0.3\arcsec\,resolution in all five disks, while H$^{13}$CO$^+$ is detected (SNR$>6 \sigma$) towards GM~Aur and HD 163296 and tentatively detected (SNR$>3 \sigma$) towards the other disks by a matched filter analysis. %(Appendix \ref{appendix:matched_filter}). 
Inside a radius of $R\sim 100$ au, the HCO$^+$ column density is flat or shows a central dip. At outer radii ($\gtrsim 100$ au), the HCO$^+$ column density decreases outwards, while the column density ratio of HCO$^+$/CO is mostly in the range of $\sim 10^{-5}-10^{-4}$. We derived the HCO$^+$ abundance in the warm CO-rich layer, where HCO$^+$ is expected to be the dominant molecular ion.
%using the column density ratio of HCO$^+$/CO and the CO depletion factor obtained in MAPS V. 
At $R\gtrsim 100$ au, the HCO$^+$ abundance is $\sim 3 \times 10^{-11} - 3\times 10^{-10}$, which is consistent with a template disk model with X-ray ionization. At the smaller radii, the abundance decreases inwards,
%at $\lesssim 100$ au in the disk of IM Lup, HD 163296, and MWC 480, and at $\lesssim 50$ au in GM Aur and AS 209.The lower abundance in the inner radius is consistent with the theoretical expectation that the ionization degree 
which indicates that the ionization degree is lower in denser gas, especially inside the CO snow line, where the CO-rich layer is in the midplane. Comparison of template disk models with the column densities of HCO$^+$, N$_2$H$^+$, and N$_2$D$^+$ indicates that the midplane ionization rate is $\gtrsim 10^{-18}$\,s$^{-1}$ for the disks around IM Lup, AS 209, and HD 163296.
%which would be caused by the CO sublimation in the midplane inside the CO snow line, where the ionization degree is low.
%is also consistent with the theoretical expectation; the ionization degree 
%is lower in denser gas, especially inside the CO snow line, where the CO-rich layer extends downwards to the midplane.
We also find hints of an increased HCO$^+$ abundance around the location of dust continuum gaps in AS~209, HD~163296, and MWC 480.
%We also find a hint of a correlation between the HCO$^+$ abundance and a gas gap carved by planets around AS 209 and HD 163296.
%A disk chemical model with the midplane ionization rate of $\zeta=1.0\times 10^{-18}$ s$^{-1}$) and the X-ray ionization dominating in the upper layers is in agreement with the HCO$^+$ and N$_2$H$^+$ column densities obtained in our analysis and MAPS X.
%A template disk chemical model in which ionization is provided by X-rays and the decay of short-lived radioactive nuclei ($\zeta_{\rm SLR}=1.0\times 10^{-18}$ s$^{-1}$) is in good agreement with our HCO$^+$ column densities, as well as \ce{N2H+} column densities derived by an accompanying paper.
\textbf{This paper is part of the MAPS special issue of the Astrophysical Journal Supplement.}
\end{abstract}


\keywords{Astrochemistry --- Exoplanet formation --- Interferometry ---  Millimeter astronomy --- Protoplanetary disks}


\section{Introduction} \label{sec:intro}

%The Molecules with ALMA at Planet-forming Scales (MAPS) Project \citep{oberg20} observed more than 20 molecular lines in Band 3 and Band 6 in five protoplanetary disks:  IM Lup, GM Aur, AS 209, HD~163296, and MWC 480. Here, we analyze the $J=1-0$ transition of HCO$^+$ (and H$^{13}$CO), which is considered to be the major molecular ion in the disks.
%Protoplanetary disks are partially ionized. 
Protoplanetary disks are the birth site of planetary systems including our Solar System. The rate and degree of ionization are important parameters for both the physical and chemical evolution of the disk, and thus for the formation of planetary systems. The disk gas is much denser, and thus its ionization degree $x_{\rm i}$, which is the relative abundance of electrons to H$_2$, is much lower than in parental molecular clouds \citep[e.g.][]{umebayashi88}\footnote{Considering neutrality, the electron abundance should be the same as the total abundance of cations. In dense midplane regions, grain particles can be the dominant charge carrier.}. Yet the ionization degree is expected to be sufficient for the gas to be partially coupled with the magnetic fields present in the disk, which could induce magneto-hydrodynamic instabilities, disk winds, and thus the angular momentum transfer needed for mass accretion \cite[e.g.][]{suzuki09, bai13}. More specifically, the ionization degree, and thus the coupling between gas and magnetic fields (i.e.\ non-ideal MHD effects) should vary spatially within the disk, which affects the physical structure of the disk \citep[e.g.][]{bethune17}.
Ionization also triggers ion-molecule reactions in the disk. For example, carbon monoxide, which is the main tracer of disk gas, can be converted to other species by ion-molecule reactions within the typical evolutionary timescale of the disk ($10^6$ yr), if the ionization rate $\zeta$ is similar to or higher than the cosmic-ray ionization rate in the ISM \citep[$10^{-17}$ s$^{-1}$;][]{furuya14,bergin14, bosman18, schwarz18}.

The ionization degree is determined by the ionization rate, gas density, and number density and size distribution of dust grains. For protoplanetary disks, there are several possible ionization sources: X-rays from the central star, cosmic rays (CRs) and stellar energetic particles, and the decay of short-lived radio active nuclei (SLR) \citep[e.g.][]{umebayashi81, glassgold97, umebayashi09, rab17}. X-rays are the dominant ionization source in the disk surface. However, X-rays are significantly attenuated before reaching the disk midplane. The attenuation length corresponds to a hydrogen column density of $N_{\rm H}\sim 10^{22}$ cm$^{-2}$ at 1 keV, while the hydrogen column densities of our target disks are $\sim 10^{23}-10^{25}$ cm$^{-2}$ at the radius of 100 au \citep{zhang20}. Although the attenuation length is larger at higher energies, the X-ray ionization rate is still expected to be $\lesssim 10^{-18}$ s$^{-1}$ in the midplane (e.g.\ \S \ref{sec:discussion})
\citep[see also][]{rab18}.
%which is much smaller than the disk gas column density in the Minimum Mass Solar Nebula: $N_{\rm H}\sim 7\times 10^{26}\times (\frac{R}{1\,{\rm au}})^{-3/2}$ cm$^{-2}$ or $\Sigma_{\rm gas}=1.7\times 10^3 \times (\frac{R}{1\,{\rm au}})^{-3/2}$ g cm$^{-2}$ \citep{hayashi81}. 
%While current disk observations suggest that the power-law index of the disk gas surface density is shallower \citep[e.g.\ $\Sigma_{\rm H} \propto R^{-1}$,][]{miotello14,zhang20}, it does not change the conclusion that X-ray ionization will be significantly attenuated before reaching the disk midplane.
The cosmic-ray ionization rate is $\sim 5 \times 10^{-17}$ s$^{-1}$ in molecular clouds \citep{dalgarno06}, and its attenuation length is much larger than that of X-rays. While \citet{umebayashi81} evaluated the attenuation length to be 96 g cm$^{-2}$, more recent work by \citet{padovani18} reassessed the propagation of CR particles and obtained an even larger attenuation length. But CRs can be scattered by magnetized stellar winds and/or magnetic fields in the disk \citep[e.g.][]{umebayashi81, cleeves14, padovani18}. Stellar energetic particles could also be disturbed by magnetic fields. Finally, the ionization rate from the decay of $^{26}$Al is estimated to be $10^{-18}$ s$^{-1}$ in the primordial Solar System based on meteorite analysis \citep[e.g.][]{umebayashi09}. While the abundance of SLRs should vary among star-forming regions, recent chemo-hydrodynamic simulations of the Milky Way Galaxy predict that the abundance of $^{26}$Al in the primordial Solar System may be typical for other star-forming regions \citep{fujimoto18}. In summary, the dominant ionization source and ionization degree should vary spatially within the disk and could be different between disks.

\begin{figure}
\epsscale{0.5}
\plotone{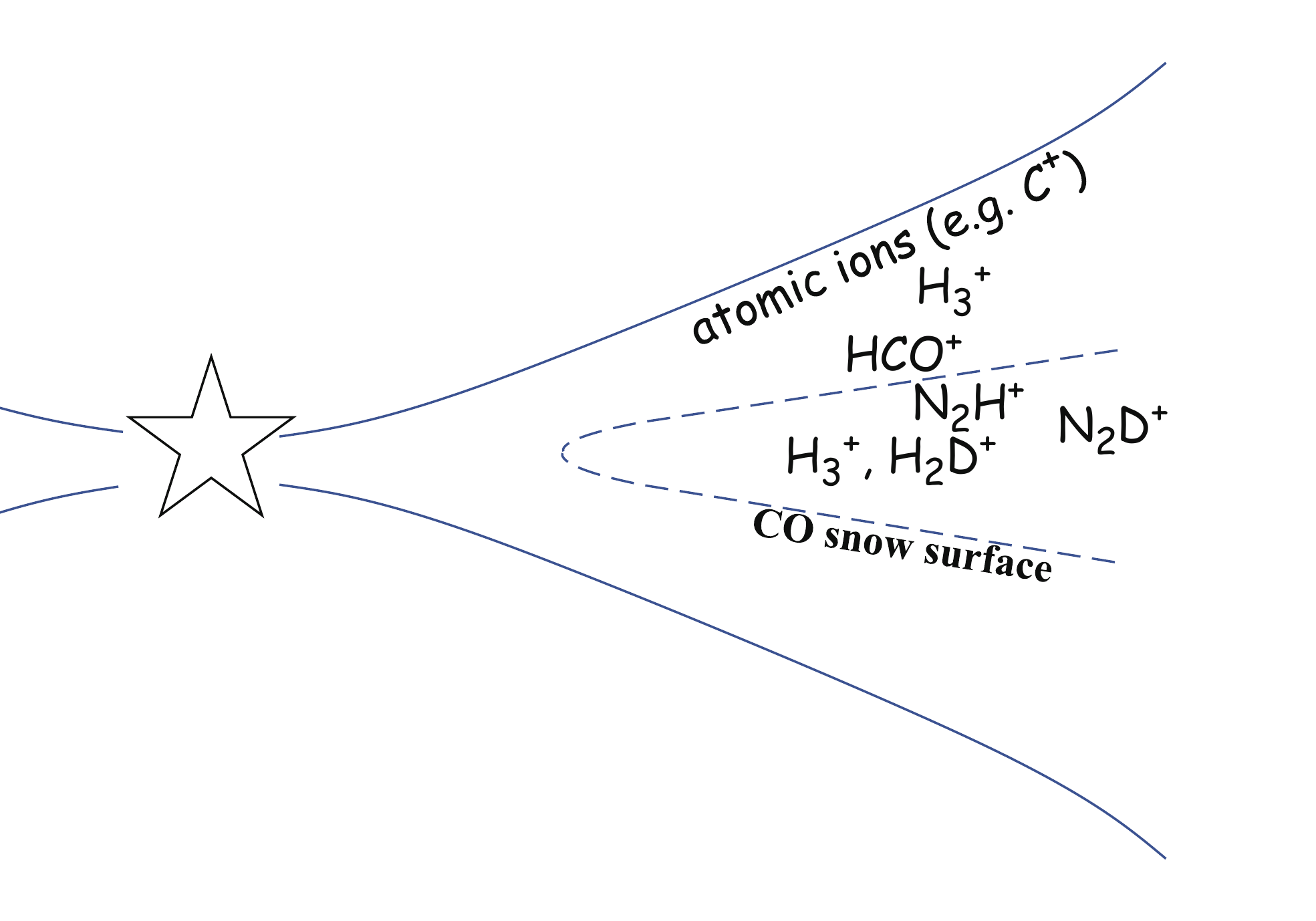}
\caption{Schematic distribution of major ions in protoplanetary disks.\label{fig:ionization_scheme}}
\end{figure}

The importance of understanding ionization processes motivates the observation of molecular ions in protoplanetary disks. Theoretical models show that the major molecular ions in disks are H$_3^+$, HCO$^+$, N$_2$H$^+$, and their deuterated isotopologues \citep[e.g.][]{aikawa01, bergin07, willacy07, aikawa15, aikawa18}. The most abundant molecular ion varies both radially and vertically in disks (Figure \ref{fig:ionization_scheme}). The disk surface is the Photon Dominated Region (PDR); photoionization makes atomic ions such as C$^+$ and S$^+$ the main charge carriers. Deeper in the disk, H$_3^+$ becomes the most abundant ion, while HCO$^+$ becomes dominant when the abundance ratio $n$(CO)/$n$(e) is higher than $\sim 10^3$, where $n$($i$) denotes the number density of species $i$ \citep{aikawa15}. In the regions where $T<20$\,K (i.e.\ below the CO snow surface and outside the CO snow line), CO freezes out onto grains, which enhances the N$_2$H$^+$ abundance. As temperature declines towards deeper layers, the dominant ion changes from N$_2$H$^+$ to H$_3^+$ and its deuterated isotopologues. Unfortunately, H$_3^+$ cannot be observed at millimeter wavelengths, and its deuterated isotopologues have not been detected in disks so far. N$_2$D$^+$ is considered to be an alternative probe to constrain the ionization degree in the cold midplane \citep{cleeves14}.
%Except for H$_3^+$, these molecular ions can be observed at mm wavelengths.
%In the MAPS collaboration, the present work analyse HCO$^+$, while the analysis of N$_2$D$^+$, together with the archival data of N$_2$H$^+$, are reported by \citet{cataldi20}.

While the rotational transitions of HCO$^+$, N$_2$H$^+$, and their isotopologues have been observed towards several disks, a quantitative evaluation of their column densities and the disk ionization degree has been reported only in a limited number of references.
\citet{oberg11a} used the IRAM 30 m telescope to observe H$^{13}$CO$^+$ $J=3-2$ in the disk of DM~Tau. They combined their observation with previous Submillimeter Array (SMA) data of N$_2$H$^+$ $J = 3-2$, HCO$^+$ $J = 3-2$, and DCO$^+$ $J = 3-2$ to estimate the ionization degree in three temperature regions of the disk: in the upper warm molecular layer ($T >20$ K), the ionization degree $x_{\rm i}$ is estimated to be $4\times 10^{-10}$ based on the HCO$^+$ data. In the cooler molecular layer ($T = 16-20$ K), where N$_2$H$^+$ and DCO$^+$ would be abundant, $x_{\rm i}$ is derived to be $3\times 10^{-11}$, while in the cold, dense midplane ($T < 16$ K), the non-detection of H$_2$D$^+$ constrains $x_{\rm i}$ to be $< 3 \times 10^{-10}$.
\citet{teague15} observed HCO$^+$ $J=1-0$ and $J=3-2$, and DCO$^+$ $J=3-2$ in DM~Tau at $\sim 1.5\arcsec$ resolution using the Plateau de Bure Interferometer (PdBI). The column densities of HCO$^+$ and DCO$^+$ at a radius of 100 au are derived to be $9.8 \times 10^{12}$ and $1.2 \times 10^{12}$ cm$^{-2}$, respectively.
They derive an ionization degree of $\sim 10^{-7}$ from the abundance ratio of DCO$^+$/HCO$^+$, assuming steady state balance between
H$_3^+$ + HD $\rightarrow$ H$_2$D$^+$ + H$_2$,
H$_2$D$^+$ + CO $\rightarrow$ DCO$^+$ + H$_2$,
and the destruction of DCO$^+$ 
%via recombination and proton transfer to CO
\citep{caselli02}. \cite{cleeves15} calculated chemical models of the disk around TW~Hya for various cosmic-ray ionization rates $\zeta_{\rm CR}$ and X-ray spectra. Instead of evaluating molecular column densities from the observational data, they calculated the disk-integrated flux of molecular lines from the models to compare with their own HCO$^+$ and H$^{13}$CO$^+$ $J=3-2$ observations of TW~Hya, and HCO$^+$, H$^{13}$CO$^+$, and N$_2$H$^+$ data from the literature \citep[e.g.][]{qi13a, qi13b}. They concluded that the observations are best reproduced by the model with low CR ionization rate $\zeta_{\rm CR}\lesssim 10^{-19}$ s$^{-1}$ and modest X-ray spectra, in which the ionization degree is $\sim 10^{-11}-10^{-10}$ near the midplane outside of a radius of $\sim 100$ au. Since the ionization degree varies spatially and between disks, observations with higher spatial resolution and towards more targets are highly desirable.

In light of the ring-gap structures observed in the millimeter dust continuum of many disks in recent years \citep[e.g.][]{andrews18, huang18, cieza21}, it is also interesting to investigate whether and how the molecular ion abundances and ionization degree correlate with the dust substructures. Theoretical studies predict that HCO$^+$ is sensitive to, and thus could be a good probe of, gas density and/or small dust abundance. In the region where HCO$^+$ is the most abundant positive charge carrier, its abundance (i.e.\ ionization degree) should be proportional to $n_{\rm H}^{-1/2}$ \citep[e.g.][]{duley84}. In the gap region, the ionization degree could also be enhanced due to deeper penetration of X-rays. \cite{aikawa06}, on the other hand, showed that the HCO$^+$ column density declines as the abundance of sub-micron sized dust grains decreases with grain growth. This is because the PDR, in which atomic ions dominate over molecular ions, extends deeper into the disk when the abundance of sub-micron dust grains decreases.
Similar results are obtained in models of disks around Herbig Ae stars \citep{jonkheid07} \cite[see also][]{wakelam19}. While these models do not include any ring-gap structures, they indicate that the HCO$^+$ abundance could vary over the substructures. The HCO$^+$ abundance would be higher in the gap than in the ring, if the gas density is lower but the UV attenuation is sufficient in the gap. On the other hand, the HCO$^+$ abundance would be lower in the gap if the UV penetration is significant \citep[see also][]{smirnov20}.
%if the gas density is lower in the gap, while it would be lower if the small dust grains are less abundant in the gap 

In the present work, we analyze the $J=1-0$ transition of HCO$^+$ and H$^{13}$CO$^+$ towards five disks around IM~Lup, GM~Aur, AS~209, HD~163296, and MWC~480 observed as a part of the Molecules with ALMA at Planet-forming Scales (MAPS) Large Program \citep[][project code 2018.1.01055.L]{oberg20}. For a given column density of HCO$^+$ at a lukewarm temperature (a few tens of K), which corresponds to the CO rich molecular layer in the disk, the $J=1-0$ line tends to be optically thinner than higher transitions, and thus useful to derive the column density. 
Our spatial resolution, 0.3$\arcsec$, is high enough to investigate the radial distribution of HCO$^+$ and to marginally resolve the continuum gap regions. We also combine our results with the analysis of CO isotopologues, N$_2$H$^+$, and N$_2$D$^+$ in the companion MAPS papers by \citet{zhang20} and \citet{cataldi20} to estimate the radial variation in the ionization degree and the ionization rate in the disks.
Various data products of the MAPS project, including the present work, such as reduced observational data, zeroth-moment maps, as well as derived HCO$^+$ column densities can be downloaded at \url{www.alma-maps.info}.

The outline of this paper is as follows. We briefly describe the observations in \S \ref{sec:observations}. \S \ref{sec:observational_results} describes the observational results: the zeroth moment maps, radial emission profiles, azimuthally averaged spectra, and molecular column densities. We derive the HCO$^+$ abundance, which corresponds to a lower limit of the ionization degree in the warm molecular layer, and compare our results with a template disk chemistry model in \S \ref{sec:discussion}. Our conclusions are presented in \S \ref{sec:conclusions}.

%Booth+19
%HD97048: enhancement of H13CO+/HCO+
%H13CO+ not detected towards HD100546

\section{Observations}\label{sec:observations}
The observational setup chosen by MAPS covered four spectral setups: two in Band 3 at $\sim$3\,mm and two in Band 6 at $\sim$1\,mm. The $J=1-0$ transitions of HCO$^+$ and H$^{13}$CO$^+$ are in one of the Band 3 setups. The spectral resolution is 0.237\,km\,s$^{-1}$ for HCO$^+$ and 0.488\,km\,s$^{-1}$ for H$^{13}$CO$^+$. Molecular data for the targeted lines is taken from the CDMS database\footnote{\url{https://cdms.astro.uni-koeln.de/}} \citep{muller01, endres16}  
%LAMDA database\footnote{\url{https://home.strw.leidenuniv.nl/~moldata/}} \citep{Schoier05} 
and summarized in table \ref{tab:molecular_data}.

Two array configurations were used: a short baseline configuration that was sensitive to extended structure, and a long baseline configuration to achieve high angular resolution. The details of the data calibration are described in \citet{oberg20}. The MAPS collaboration produced images from the calibrated visibilities using the CLEAN algorithm implemented in the CASA \texttt{tclean} task. Details of the imaging procedure are described in \citet{czekala20}. We use the fiducial images provided by MAPS, which have a 0\farcs3 circular beam.  This beam size corresponds to a physical size of 30.3\,au at the distance of HD~163296, which is the closest target, and 48.6\,au at the distance of MWC~480, which is the most distant. Basic properties of the data cubes such as noise levels are summarized in Table 11 and 12 in \citet{oberg20}.

%Initial calibration is performed using the ALMA pipeline. The phase centres of the independent execution blocks are aligned using the continuum peak. Then, self-calibration was performed on the short baseline data only (1--6 rounds of phase-only self-calibration, followed by 0--1 rounds of amplitude self-calibration). Finally, the self-calibrated short baseline data were combined with the long baseline data, and the combined data set was self-calibrated.

\begin{table*}
\centering
\caption{Molecular data of the targeted lines taken from the CDMS database.}\label{tab:molecular_data}
\begin{tabular}{cCCC}
\hline
transition & \text{frequency}  & A_{ij}\tablenotemark{a} & E_u\tablenotemark{b}\\
& \text{[GHz]} & [\mathrm{s}$^{-1}$] & \text{[K]} \\
\hline
\ce{HCO+} J = 1--0 & 89.1885247 & 4.2512\times10^{-5} & 4.28\\
H$^{13}$CO$^+$ J = 1--0 & 86.7542884 & 3.8534\times10^{-5} & 4.16\\
\hline
\end{tabular}
\tablenotetext{a}{Einstein A coefficient.}
\tablenotetext{b}{Upper state energy.}
\end{table*}

%\begin{table*}
%\centering
%caption{Properties of the data cubes used in this study.\textbf{Make sure this is consistent with MAPS I}}\label{tab:image_params}
%\begin{tabular}{ccccc}
%\hline
%emission line & rms  & beam & spectral resolution & channel width\\
%& [mJy\,beam$^{-1}$] & & [km\,s$^{-1}$] & [km\,s$^{-1}$] \\
%\hline\hline
%\multicolumn{5}{c}{IM Lup}\\
%HCO$^+$ 1--0 & & $0.3\arcsec\times0.3\arcsec$ & & \\
%H$^{13}$CO$^+$ 1--0 & & $0.3\arcsec\times0.3\arcsec$ & & \\
%hline
%\multicolumn{5}{c}{GM~Aur}\\
%CO$^+$ 1--0 & & $0.3\arcsec\times0.3\arcsec$  & & \\
%H$^{13}$CO$^+$ 1--0 & & $0.3\arcsec\times0.3\arcsec$ & & \\
%\hline
%\multicolumn{5}{c}{AS 209}\\
%HCO$^+$ 1--0 & & $0.3\arcsec\times0.3\arcsec$  & & \\
%H$^{13}$CO$^+$ 1--0 & & $0.3\arcsec\times0.3\arcsec$ & & \\
%\hline
%\multicolumn{5}{c}{HD~163296}\\
%
%HCO$^+$ 1--0 & & $0.3\arcsec\times0.3\arcsec$ & & \\
%H$^{13}$CO$^+$ 1--0 & & $0.3\arcsec\times0.3\arcsec$ & &\\
%\hline
%\multicolumn{5}{c}{MWC~480}\\
%
%HCO$^+$ 1--0 & & $0.3\arcsec\times0.3\arcsec$ (-) & & \\
%H$^{13}$CO$^+$ 1--0 & & $0.3\arcsec\times0.3\arcsec$ (-)& &  \\
%\hline
%\end{tabular}
%\end{table*}

\section{Observational results}\label{sec:observational_results}

\subsection{Zeroth moment maps, disk-integrated fluxes, and radial emission profiles} \label{subsec:mom0_radial_emission_profiles}

The MAPS collaboration produced zeroth moment maps by applying a Keplerian mask to the data cubes and integrating over the velocity axis \citep{law20_radial_profiles}. The mask parameters are conservatively chosen to incorporate all $^{13}$CO $J=2-1$ emission, which is the most widespread emission of any species except for CO $J=2-1$. The mask parameters for the five disk are summarized in Table 1 in \citet{czekala20}. The stellar masses adopted for the generation of the masks are listed in Table 1 in \citet{oberg20}. These maps were used for all scientific analysis, in particular to derive radial emission profiles \citep{law20_radial_profiles}. A Keplerian mask, however, introduces spatial variance in noise distribution, since the number of channels summed to create the zeroth moment maps vary spatially. The discontinuous noise distribution occasionally causes arc-like artefacts, which could be mistaken as real substructure, in the central regions. To mitigate such artefacts, MAPS also produced ``hybrid'' zeroth moment maps by combining a Keplerian mask and a smoothed $\sigma$-clip mask \citep[thus their name ``hybrid'', see][Appendix A]{law20_radial_profiles}. The latter removes the pixels with a S/N below a given threshold. We emphasize that by using a clipping mask with a threshold larger than 0$\sigma$, some emission is inevitably lost. Therefore, the hybrid zeroth moment maps are for presentational purposes only, and are not used for any quantitative analysis.

\begin{figure}
\epsscale{1.25}
\plotone{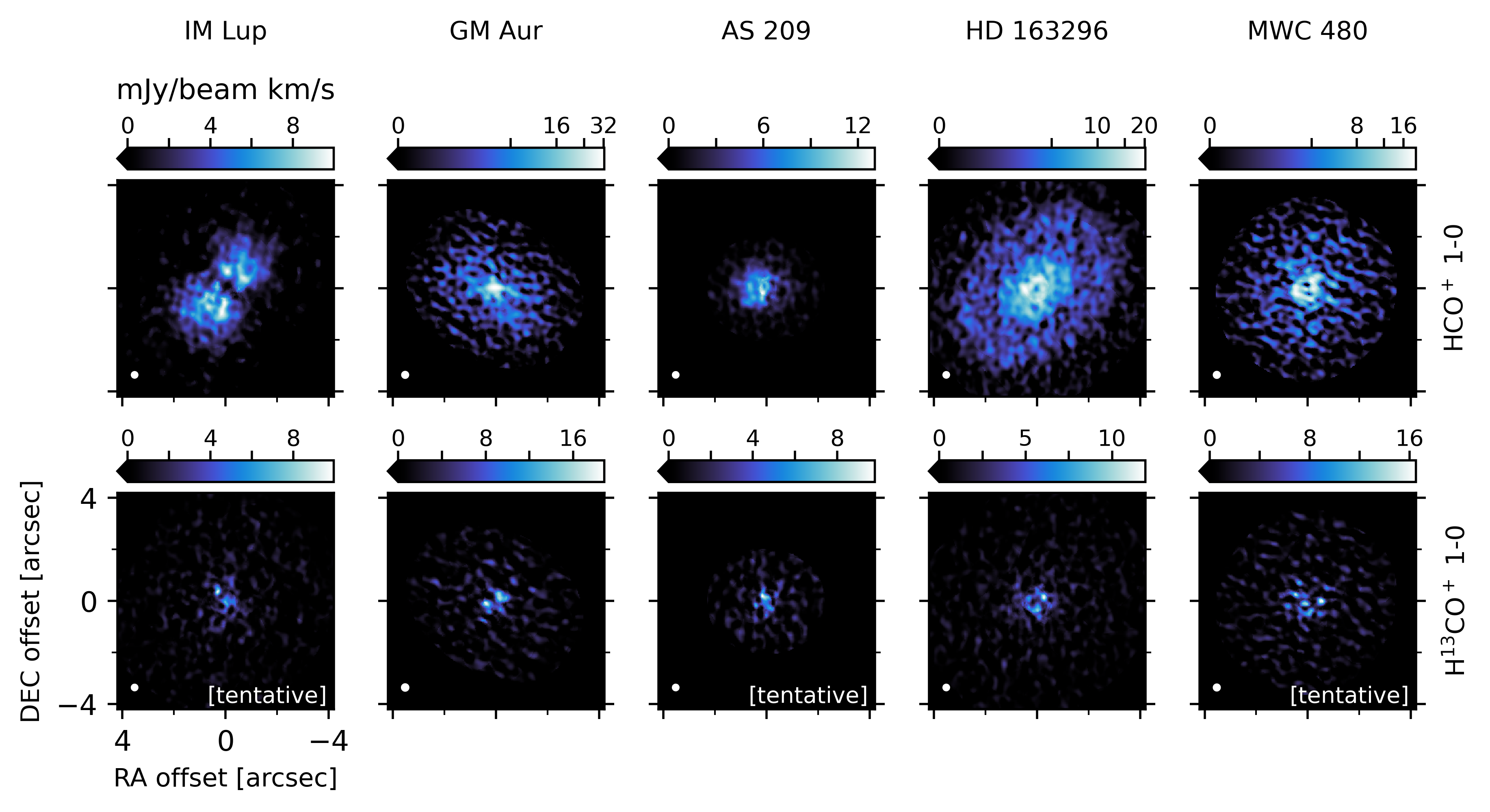}
\caption{Gallery of hybrid  zeroth moment maps for HCO$^+$ $J=1-0$ (top row) and H$^{13}$CO$^+$ $J=1-0$ (bottom row) for the MAPS sample, ordered from left to right by increasing stellar mass \citep[see][table 1]{oberg20}. These maps were generated by combining a Keplerian mask with a smoothed $\sigma$-clip mask \citep[Appendix A of][]{law20_radial_profiles}. The color scales employ either linear or arcsinh stretches, with the lower end saturating at 0\,mJy\,beam$^{-1}$\,km\,s$^{-1}$. The synthesized beam is shown by the white ellipse in the lower left of each panel. Due to the use of a mask, the noise level is not constant across a map. Lines that are only tentatively detected in total flux or matched filter are marked in the lower right of the panel.
\label{fig:HCO+_mom0_gallery}}
\end{figure}

In Figure \ref{fig:HCO+_mom0_gallery}, we show hybrid zeroth moment maps for HCO$^+$ $J=1-0$ and H$^{13}$CO$^+$ $J=1-0$. We choose the best-looking $\sigma$-clip value by visual inspection: 1$\sigma$ for \ce{HCO+} $J=1-0$ towards IM~Lup, and 0$\sigma$ for the other maps.
%It should be noted that the noise is spatially varying in these maps due to the use of masks \citep[][]{czekala20,law20_radial_profiles}. 
Hybrid zeroth moment maps with a 0\farcs5 tapered beam are shown in Figure \ref{fig:mom0_05taper} in Appendix \ref{appendix:05taper}; the faint emission is more easily seen in the lower spatial resolution images.

The HCO$^+$ $J=1-0$ line is clearly detected in all five disks. The emission is relatively bright inside a radius of $\sim 200$ au, while its diffuse emission extends to $\gtrsim 400$ au, except for the disk of AS 209, in which the emission extends only up to $R\sim 300$ au. In AS 209, the emission from the west side of the disk is absorbed by a foreground cloud, which happens to have a similar line-of-sight velocity \citep{oberg11b}. 
We thus performed all analysis for AS 209 only on the un-obscured east side of the disk within a $\pm 55$ degree wedge centered on the semi-major axis \citep{teague18b,law20_radial_profiles}.
%{\bf This part is modified referring to Charles' comment. Gianni-san, did you use the $\pm 55$ wedge? Yes}
The emission of H$^{13}$CO$^+$ $J=1-0$ is very weak and is not clearly seen in the zeroth moment maps.
%Although we see some emission-like features in Figure \ref{fig:HCO+_mom0_gallery} and Figure \ref{fig:mom0_05taper}, it is difficult to judge the detection of the line, since the noise is not uniform in the image. %The upper limit of the line flux is, however, very useful to constrain the column density of HCO$^+$, combined with HCO$^+$ $1-0$ (see \S \ref{section:column}).
%It is, however, very useful to constrain the column density of HCO$^+$ (see \S \ref{section:column}).

\begin{figure}
\epsscale{1.15}
\plotone{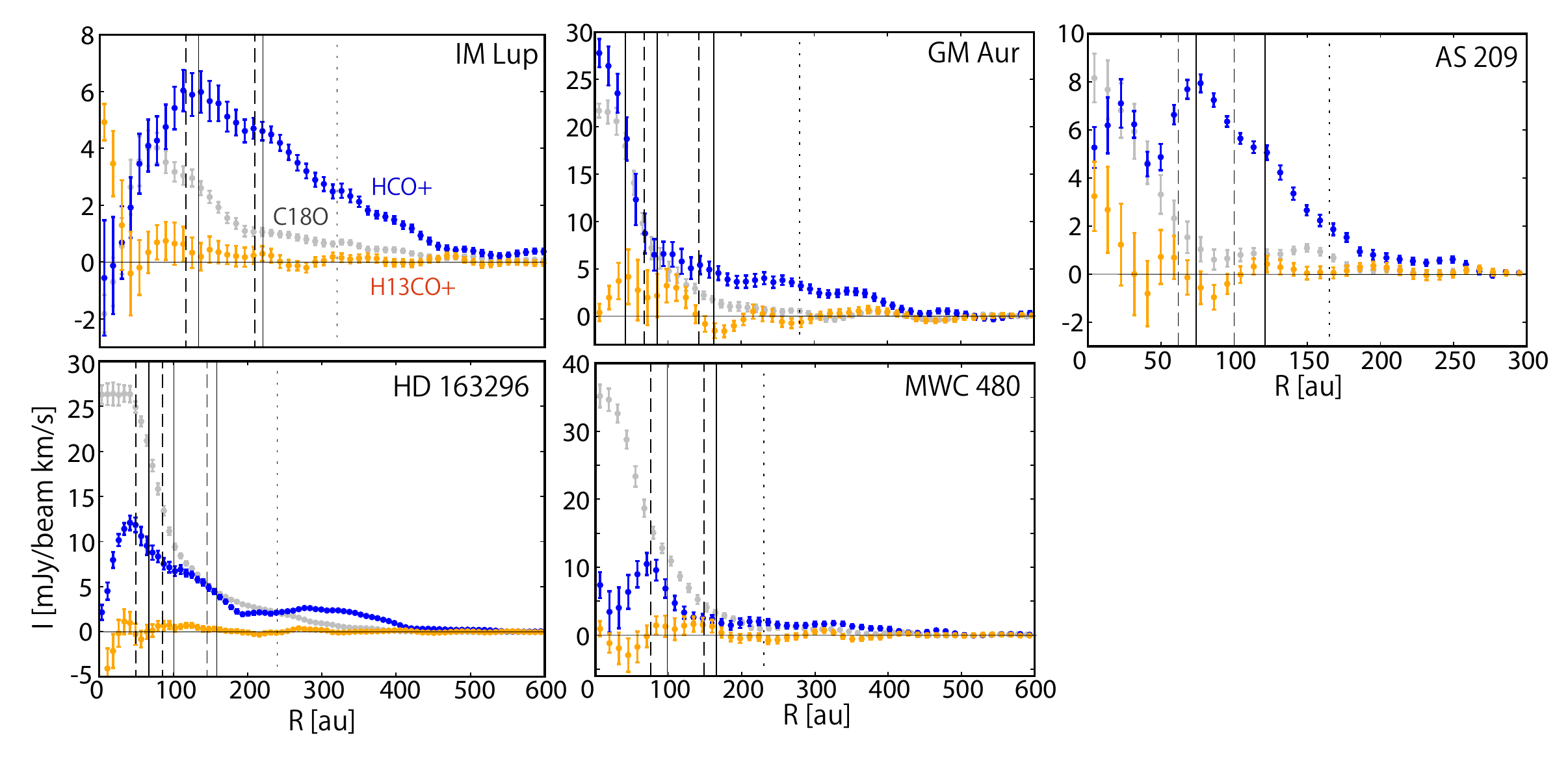}
\caption{Radial emission profiles of HCO$^+$ $J=1-0$, H$^{13}$CO$^+$ $J=1-0$, and C$^{18}$O $J=1-0$ for IM~Lup, GM~Aur, AS~209, HD~163296, and MWC~480. The error bar shows the $\pm1\sigma$ error. The size of the radial bins is 0.075\arcsec, which is a quarter of the beam size. The vertical lines indicate the positions of rings (solid), gaps (dashed), and the edge of millimeter dust continuum (dotted) referring to \citet{huang18,long18, liu19,law20_radial_profiles} \citep[see also][]{sierra20}. 
%{\bf Update the dust continuum gap and ring referring to Table 5 of Charles IIb.--DONE}
\label{fig:radial_profile}}
\end{figure}

Disk-integrated fluxes of HCO$^+$ $J=1-0$ and H$^{13}$CO$^+$ $J=1-0$ are presented in Table \ref{tab:disk_integrated_fluxes}. These were calculated by integrating the flux within a Keplerian mask. The radial extent of the Keplerian mask is different from that used for the zeroth moment maps. For \ce{HCO+}, the radial extent of the mask is visually determined from the extent of the emission seen in the radial emission profile (see Figure \ref{fig:radial_profile} and Table \ref{tab:disk_integrated_fluxes}). For H$^{13}$CO$^+$, we measured fluxes by using the same radial extents as for \ce{HCO+}. In an attempt to increase the SNR, we computed additional fluxes by adopting the same radial extent as the Keplerian models that maximize the SNR in the matched filter analysis in the $uv$ plane (Appendix \ref{appendix:matched_filter}, Figure \ref{fig:matched_filter}), as shown in the right column of Table \ref{tab:disk_integrated_fluxes}. Errors were calculated by repeating the flux measurement procedure at off-source positions and taking the standard deviation of the off-source fluxes. A 10\% flux calibration error was added in quadrature. If the resulting SNR was smaller than 3, Table \ref{tab:disk_integrated_fluxes} reports the 3$\sigma$ upper limit. To calculate these disk-integrated fluxes and errors, we used the image cubes that are not corrected for the primary beam. This is because for primary beam corrected images, the noise increases towards the edges of the image, making our approach to estimate the error from off-source positions invalid. However, we verified that the difference to fluxes extracted from primary beam corrected images is negligible. For deriving radially resolved column density profiles, we use the primary beam corrected images.

H$^{13}$CO$^+$ is seen with SNR$>3$ for IM~Lup, GM~Aur, and HD~163296, if we use the masks with the extent informed by the matched filter analysis. The matched filter analysis itself (Appendix \ref{appendix:matched_filter}) gives a response larger than 6$\sigma$ for GM~Aur and HD~163296, and larger than 3$\sigma$ for the other disks. Thus, we consider H$^{13}$CO$^+$ $J=1-0$ detected in GM~Aur and HD~163296, and tentatively detected for the other three disks.
We note that the values in Table \ref{tab:disk_integrated_fluxes} are different from those in Table 12 in \citet{oberg20}, who adopted the same Keplerian masks to generate zeroth moment maps and to estimate disk-integrated fluxes.

\begin{table*}
%\centering
\caption{Disk-integrated fluxes. Upper limits are at 3$\sigma$ significance.}\label{tab:disk_integrated_fluxes}
\begin{tabular}{c|CCCCCCCCC}
\hline
& \multicolumn{3}{c}{HCO$^+$ 1--0} & \multicolumn{3}{c}{H$^{13}$CO$^+$ 1--0\tablenotemark{a}} & \multicolumn{3}{c}{H$^{13}$CO$^+$ 1--0\tablenotemark{b}}\\
& r_\mathrm{min}\tablenotemark{c} & r_\mathrm{max}\tablenotemark{d} & \text{flux} & r_\mathrm{min} & r_\mathrm{max} & \text{flux} & r_\mathrm{min} & r_\mathrm{max} & \text{flux}\\
& \text{[au]} & \text{[au]} & \mathrm{[mJy\,km\,s^{-1}]} & \text{[au]} & \text{[au]} & \mathrm{[mJy\,km\,s^{-1}]} & \text{[au]} & \text{[au]} & \mathrm{[mJy\,km\,s^{-1}]} \\
\hline
IM Lup & 0 & 700 & 519 \pm 53 & 0 & 700 & <46 & 0 & 400 & 23 \pm 6\\
GM Aur & 0 & 500 & 412 \pm 42 & 0 & 500 & <43 & 0 & 400 & 27 \pm 7\\
AS 209 & 0 & 300 & 178 \pm 19 & 0 & 300 & <33 & 0 & 200 & <26\\
HD 163296 & 0 & 550 & 1002 \pm 101 & 0 & 550 & <54 & 50 & 400 & 27 \pm 7\\
MWC 480 & 0 & 550 & 323 \pm 34 & 0 & 550 & <34 & 50 & 100 & <18\\
\hline
\end{tabular}
\tablenotetext{a}{For a Keplerian mask with the same radial extent as used for \ce{HCO+} 1--0.}
\tablenotetext{b}{For a Keplerian mask with the same radial extent as the matched filter that maximises the SNR of the filter response (see Appendix \ref{appendix:matched_filter} and Fig.\ \ref{fig:matched_filter}).}
\tablenotetext{c}{Minimum radius of the Keplerian mask.}
\tablenotetext{d}{Maximum radius of the Keplerian mask.}
\end{table*}

Figure \ref{fig:radial_profile} shows the radial emission profiles of the $J=1-0$ transitions of HCO$^+$ and H$^{13}$CO$^+$. These profiles are produced by azimuthal averaging the zeroth moment maps that were produced with a Keplerian mask only (i.e.\ without a $\sigma$-clip mask). The inclination and position angles of each disk are listed in Table 1 in \citet{oberg20}.
%HCO$^+$ 1--0 is clearly detected for all five sources, while the integrated intensity of H$^{13}$CO$^+$ 1--0 is less than 3 sigma.
Since HCO$^+$ is expected to be formed by CO + H$_3^+$ and to be a dominant molecular ion in CO gas rich layers, we also plotted the radial emission profile of C$^{18}$O $J=1-0$ derived by \citet{law20_radial_profiles}. The vertical lines indicate the radius of rings (solid), gaps (dashed), and the dust disk edge (dotted) observed in the millimeter dust continuum \citep{huang18, long18,liu19, sierra20, law20_radial_profiles}.
Both HCO$^+$ and C$^{18}$O emission extend out to several 100 au. In the IM~Lup disk, the radial emission profile of HCO$^+$ $J=1-0$ is similar to that of C$^{18}$O, showing a ring-like distribution, although the peak position of the HCO$^+$ emission is $\sim 36$ au outside of the C$^{18}$O peak. In GM Aur, on the other hand, both the C$^{18}$O and HCO$^+$ emission are centrally peaked.
The radial emission profiles of C$^{18}$O and HCO$^+$ are quite different in AS 209: around a radius of $\sim 77$ au, the HCO$^+$ emission has a local maximum, while the radial gradient of C$^{18}$O changes from negative to plateau. Interestingly, the peak position of the HCO$^+$ emission coincides with that of a dust continuum ring. In the disks around the Herbig Ae stars, HD~163296 and MWC~480, the HCO$^+$ emission shows a central dip, while C$^{18}$O is flat for HD~163296 and centrally peaked for MWC~480.

The intensity ratio of the C$^{18}$O and HCO$^{+}$ lines varies significantly with radius and among objects. For example, the ratio reaches $\sim 5$ towards the center of the MWC~480 disk, while HCO$^+$ is brighter than C$^{18}$O around $R \sim 77$ au in AS~209. Although this indicates that the abundance (column density) ratio of HCO$^+$ to C$^{18}$O also varies, we need to consider the line optical depth. Under LTE conditions with an excitation temperature of 30 K, which is a typical temperature in the warm molecular layers, the optical depth of C$^{18}$O $J=1-0$ reaches unity for a C$^{16}$O column density of $\sim 6\times 10^{18}$ cm$^{-2}$, %$3 \times 10^{18}$ cm$^{-2}$,
assuming a $^{16}$O/$^{18}$O isotope ratio of 557 \citep{wilson99}. \citet{zhang20} derived the radial column density distribution of CO by analyzing several transitions of CO and its isotopologues. The CO column density exceeds %$3 \times 10^{18}$ cm$^{-2}$
$6\times 10^{18}$ cm$^{-2}$ 
% AS 209 166au, GM Aur 200 au, IM Lup 280 au, HD163296 314au, MWC480 336 au
inside a radius of $\sim 100$ au in the GM Aur, HD 163296, and MWC 480 disks. At least for these inner radii, C$^{18}$O $J=1-0$ is optically thick, and thus the intensity ratio of C$^{18}$O to HCO$^+$ does not reflect their column density ratio. We also need to evaluate the optical depth of HCO$^+$ to derive its column density; the optical depth of HCO$^+$ $J=1-0$ reaches unity for a HCO$^+$ column density of $\sim 1\times 10^{13}$ cm$^{-2}$ under LTE conditions at a temperature of 30 K.
%dust opacity
%can we see the backside of the disk in HCO+ in optically thick regon?

\subsection{Azimuthally averaged spectra and radial profiles of the HCO$^+$ column density} \label{section:column}

\begin{figure}
\plotone{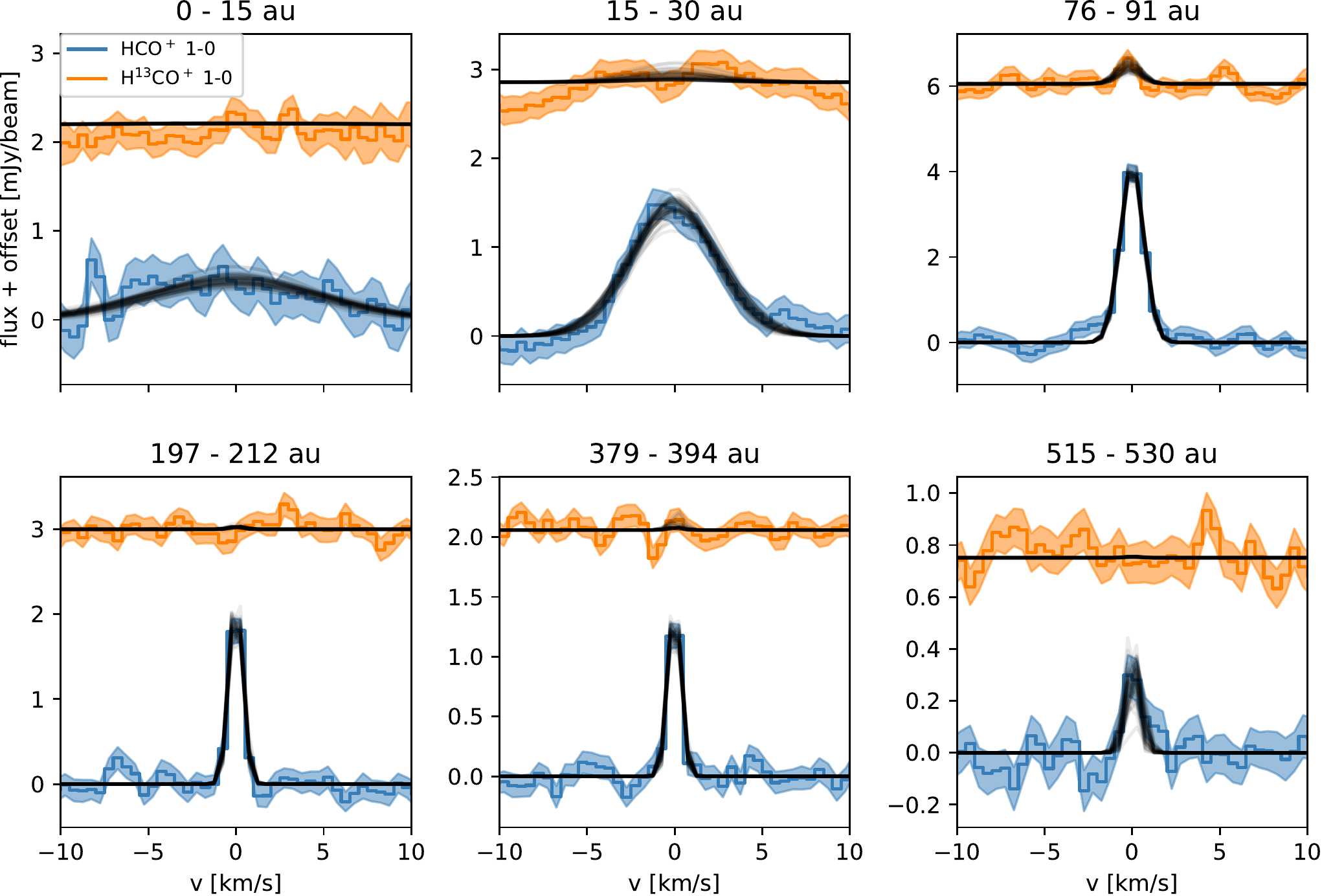}
\caption{Examples of HCO$^+$ and H$^{13}$CO$^+$ $J=1-0$ model spectra fit to azimuthally averaged spectra of HD~163296 for a few radial bins. The orange and blue solid lines show the data, with the shaded regions corresponding to the 1$\sigma$ uncertainty. The black curves show 50 randomly selected models drawn from the Monte Carlo chain, with the selection probability being proportional to the posterior probability of the model. Spectra are vertically offset for clarity.\label{fig:averaged_spectra_fits}}
\end{figure}

In order to evaluate the radial profile of the optical depth and column density of HCO$^+$, we first derive the azimuthally averaged spectrum for radial bins with a width of half the beam size. Due to the Keplerian rotation of the disk, at each spatial location in the data cube, the spectrum is shifted with respect to the systemic velocity. We thus shift each spectrum by the Keplerian velocity projected to the line of sight before averaging the spectra azimuthally. This results in all individual spectra being centered at the systemic velocity, which increases the S/N ratio of the azimuthally averaged spectrum \citep{yen16, teague16, matra17}. The calculation of the error bars of the averaged spectrum is described in detail in Appendix D of \citet{cataldi20}. Briefly, we calculate both the standard deviation of the spectrum in regions without line emission, and an analytical error bar based on the number of independent samples included in the average. We then adopt the maximum of the two as our final error bar. Figure \ref{fig:averaged_spectra_fits} shows examples of extracted spectra of HD 163296. The full gallery of spectra is found in Appendix \ref{appendix:shifted_stacked_spectra}.

We then calculate the column density of HCO$^+$, $N$(HCO$^+$), at each radial bin by simultaneously fitting the azimuthally averaged spectra of HCO$^+$ and H$^{13}$CO$^+$. We employ the same fitting procedure as \citet{cataldi20}. Briefly, for a given \ce{HCO+} column density, we compute \ce{HCO+} $J=1-0$ and H$^{13}$CO$^+$ $J=1-0$ model spectra that can be compared to the data. There are five free parameters: the logarithm of the \ce{HCO+} column density, the excitation temperature, an offset $\Delta v$ of the line center with respect to the systemic velocity for each line, and the full width at half maximum (FWHM) of the Gaussian kernel with which the model spectra are convolved to mimic observational broadening. The H$^{13}$CO$^+$ column density is fixed to 1/68 of the \ce{HCO+} column density \citep{milam05}. 
%Since HCO$^+$ is expected to be the major molecular ion in the CO gas rich layer of the disk, we assumed an excitation temperature of 30 K, typical for the warm molecular layer.
We use an MCMC method implemented in the \texttt{emcee} package \citep{Foreman-Mackey13} to explore the parameter space and assume flat priors. We employ 200 walkers taking 5000 steps each, and discard the first 2500 steps from the analysis. The free parameters and the prior boundaries are listed in Table \ref{tab:fit_parameter_priors}. The full details of the fitting procedure can be found in \citet{cataldi20}. Figure \ref{fig:averaged_spectra_fits} shows examples of model spectra fitted to the HD~163296 data. The full gallery of models is shown in Appendix \ref{appendix:shifted_stacked_spectra}.
%Although the H$^{13}$CO$^+$ emission is not detected {\bf i.e. less than 2 sigma?} in these spectra, it is useful to constrain the HCO$^+$ column density.

\begin{figure}
\epsscale{1.2}
\plotone{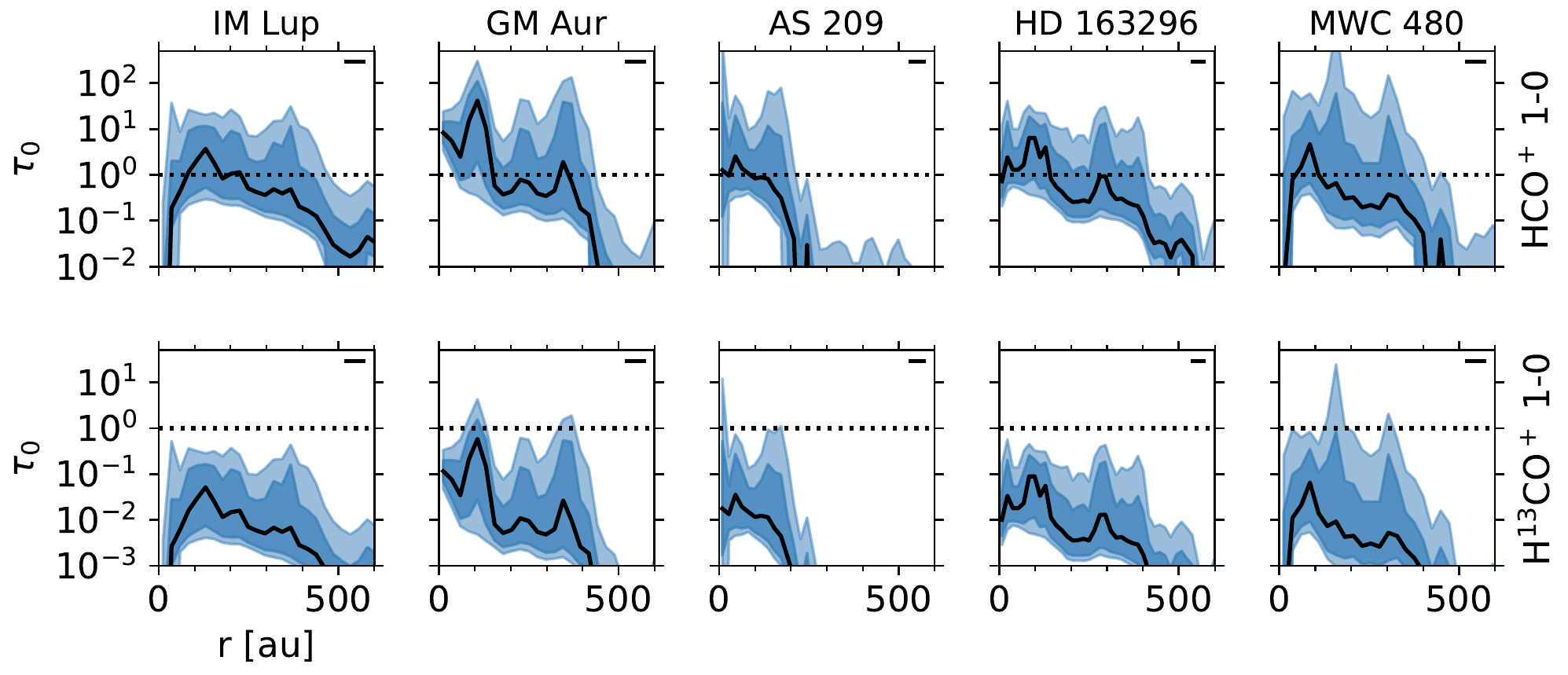}
\caption{Optical depth profiles of HCO$^+$ ($J=1-0$) and H$^{13}$CO$^{+}$ ($J=1-0$) derived from fitting azimuthally averaged spectra. The black lines show the median of the posterior probability, while the shaded regions extend between the 16th to 84th and the 2.3th to 97.7th percentiles. The beam size is shown as a horizontal line in the upper right. \label{fig:HCOP_optical_depth}}
\end{figure}

\begin{figure}
\epsscale{1.2}
\plotone{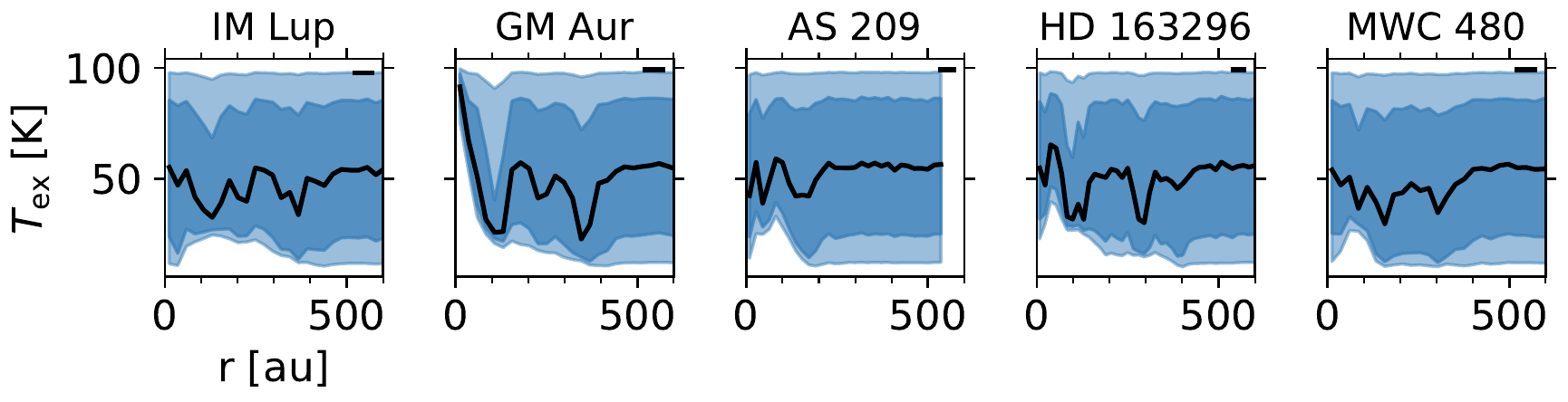}
\caption{The excitation temperature derived from fitting azimuthally averaged spectra. The black lines show the median of the posterior probability, while the shaded regions extend between the 16th to 84th and the 2.3th to 97.7th percentiles. The beam size is shown as a horizontal line in the upper right. \label{fig:HCOP_Tex}}
\end{figure}

\begin{figure}
\epsscale{1.2}
\plotone{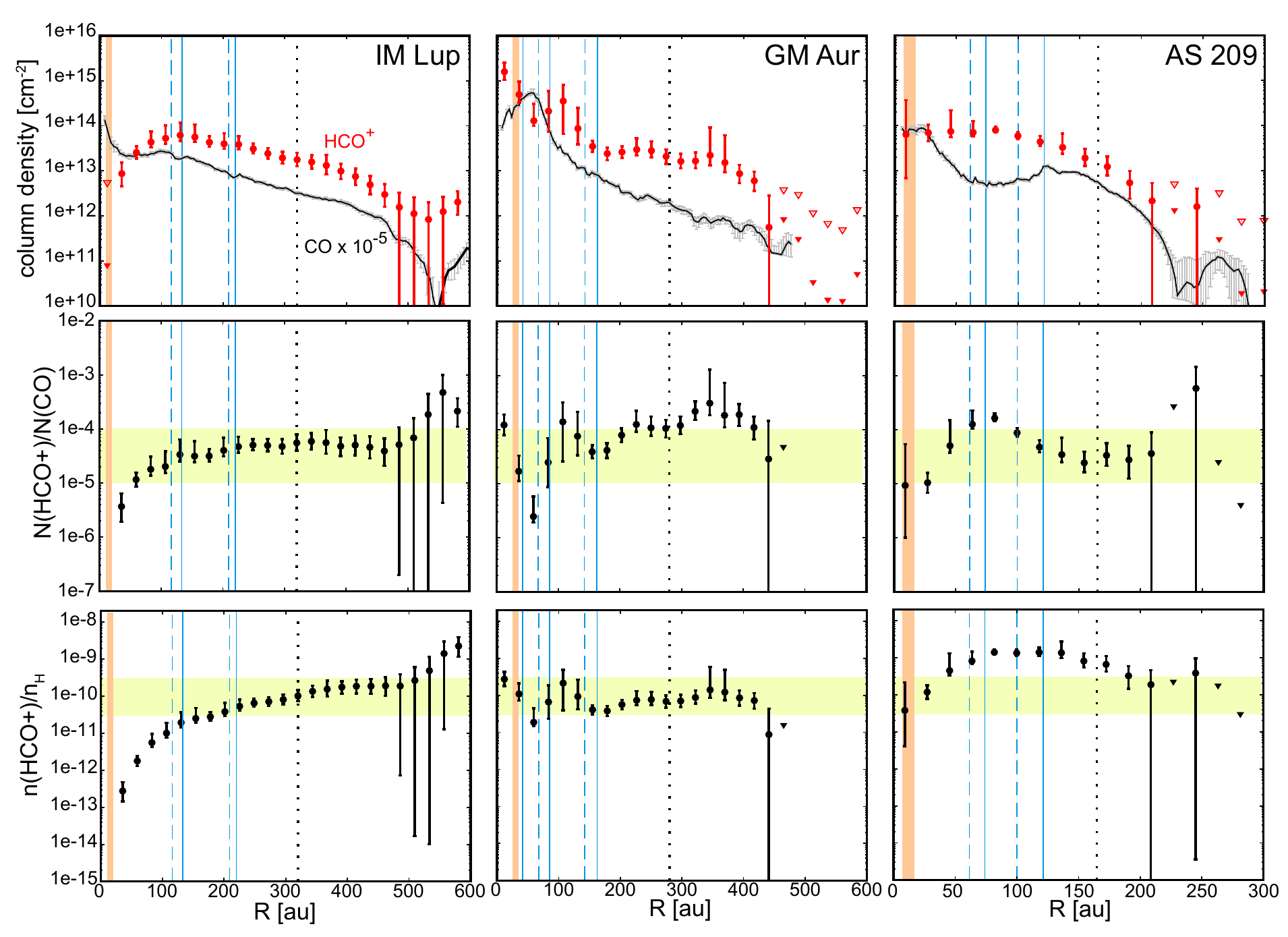}
\caption{{\em (Top)} Column densities of HCO$^+$ (red) and CO column density scaled by $10^{-5}$ (black lines with gray error bars) . The error bars correspond to  10 \% or 1 $\sigma$ for CO (see \S 3.2), while they show the 16th and 84th percentile for HCO$^+$. At radii where the median value of the molecular column density is lower than the value at 16th percentile by a factor of $>10$, we plot the upper limits as inverted triangles (84 percentile for closed triangles and 98 percentile for open triangles). The vertical orange bars mark the CO snowline as inferred from the model by \citet{zhang20} with error of $\pm 5$ au. The blue solid and dashed lines depict the radius of rings and gaps, respectively, in dust continuum. The black dotted lines depict the outer edge of the millimeter dust continuum.  {\em (Middle)} The column density ratio of HCO$^+$ to CO. The yellow bars mark the HCO$^+$/CO column density ratio of $10^{-5}-10^{-4}$. {\em (Bottom)} Abundance of HCO$^+$ relative to hydrogen nuclei in the CO gas rich layer. The yellow bars mark the abundance of $3\times 10^{-11}-3\times 10^{-10}$. In the middle and bottom panels, we consider only the error in HCO$^+$ column density, since it dominates over the error of CO column density estimates. At radii where the median value of the HCO$^+$ column density is lower than the value at 84th percentile by a factor of $>10$, we plot the value which corresponds to the 1-$\sigma$ upper limit (i.e.\ 84 percentile of HCO$^+$ column density) with an inverted triangle. 
\label{fig:col_HCOp_comb1}}
\end{figure}

Figure \ref{fig:HCOP_optical_depth} and \ref{fig:HCOP_Tex} show the optical depths and excitation temperature derived by fitting the azimuthally averaged spectra. We find that HCO$^+$ $J=1-0$ is optically thick at the column density peaks. However, we are still able to constrain the HCO$^+$ column density because the H$^{13}$CO$^+$ $J=1-0$ transition is in the optically thin regime. The top panels in Figure \ref{fig:col_HCOp_comb1} present the derived HCO$^+$ column densities.

The excitation temperature is poorly constrained for large radial regions of the disks, as shown in Fig.\ \ref{fig:HCOP_Tex}. However, at least a lower limit on the excitation temperature can be placed in some disk locations. The innermost $\sim$50\,au of GM~Aur stand out with an apparently well constrained $T_\mathrm{ex}\approx90$\,K. This region also requires a large FWHM of 17\,km\,s$^{-1}$. These rather extreme parameters might indicate that the fit is not reliable. In fact, considerable broadening is seen for all disks for the two innermost radial bins (i.e.\ within one beam FWHM from the disk center) due to beam smearing of the large velocity gradient in the inner disk (see Figures \ref{fig:IM_Lup_az_spectra}
%, \ref{fig:GM_Aur_az_spectra}, \ref{fig:AS_209_az_spectra}, \ref{fig:HD_163296_az_spectra} and \ref{fig:MWC_480_az_spectra}
). Therefore, the column density estimates for the two innermost radial bins should be considered with caution.

In Appendix \ref{appendix:HCOp_column_density_T_dependence}, we show optical depth and column density profiles for additional fits where the excitation temperature has been fixed to 30\,K, which is a typical temperature for the warm molecular layer where \ce{HCO+} is expected to be present. The corresponding models are shown in %Figures \ref{fig:IM_Lup_az_spectra} to \ref{fig:MWC_480_az_spectra}.
Figure \ref{fig:IM_Lup_az_spectra}.
The column densities generally agree well with the fits where $T_\mathrm{ex}$ is a free parameter, as can be seen in Fig.\ \ref{fig:N_HCO+_Tex30_comparison}. Furthermore, for both fits, the model spectra fit the data well
%(Figures \ref{fig:IM_Lup_az_spectra} to \ref{fig:MWC_480_az_spectra}). 
(Figure \ref{fig:IM_Lup_az_spectra}). 
One exception is the region inwards of $\sim$150\,au towards GM~Aur where the models with $T_\mathrm{ex}$ as a free parameter generally provide a better fit, especially for the two innermost radial bins discussed in the previous paragraph
%(Fig.\ \ref{fig:GM_Aur_az_spectra}). 
(Fig.\ \ref{fig:IM_Lup_az_spectra}). 
The other exception is the region inwards of 76\,au towards HD~163296, where the fits with $T_\mathrm{ex}=30$\,K underpredict the \ce{HCO+} emission and overpredict the H$^{13}$CO$^+$ emission
%(Fig.\ \ref{fig:HD_163296_az_spectra}).
(Fig.\ \ref{fig:IM_Lup_az_spectra}).
Here, the models with $T_\mathrm{ex}$ as a free parameter provide a better fit and predict a column density smaller by a factor of a few compared to the fits where $T_\mathrm{ex}=30$\,K.

\begin{table*}
%\centering
\caption{Free parameters for fitting of azimuthally averaged spectra to derive the \ce{HCO+} column density.}\label{tab:fit_parameter_priors}
\begin{tabular}{Cccc}
\hline
\mathrm{Parameter} & prior low\tablenotemark{a} & prior high\tablenotemark{b} & unit\\
\hline
\log N_\mathrm{HCO^+} & 1 & 16 & $\log(\mathrm{[cm^{-2}]})$\\
 $T_{\rm ex}$ &  10 & 100  &  [K]\\
\mathrm{FWHM_\mathrm{kernel}} & 0.5\tablenotemark{c} & variable\tablenotemark{d} & [km\,s$^{-1}$]\\
\Delta v_\mathrm{HCO^+}\tablenotemark{e} & $-$0.3 & 0.3 & [km\,s$^{-1}$]\\
\Delta v_\mathrm{H^{13}CO^+} & $-$0.3 & 0.3 & [km\,s$^{-1}$]\\
\hline
\end{tabular}
\tablenotetext{a}{Lower bound of flat prior.}
\tablenotetext{b}{Upper bound of flat prior.}
\tablenotetext{c}{Equal to the channel width.}
\tablenotetext{d}{Initial value for innermost radial bin is 20\,km\,s$^{-1}$. Dynamically adjusted when sequentially fitting larger and larger radii \citep[see][for details]{cataldi20}.}
\tablenotetext{e}{Offset of the line center with respect to the systemic velocity.}
\end{table*}

%derived assuming an excitation temperature of 30 K. We also performed an additional calculation assuming that the excitation temperature is equal to the brightness temperature of $^{13}$CO $J=2-1$ evaluated by \citet{law20_surfaces_vertical_distributions} (see Figure \ref{fig:N_HCO+_T13CO} in Appendix \ref{appendix:HCOp_column_density_T_dependence}). The derived HCO$^+$ column densities are similar between these two cases.
In Figure \ref{fig:col_HCOp_comb1}, we also plot the column densities of CO which are derived from C$^{18}$O $J=2-1$ emission in \citet{zhang20}.
For each disk, they constructed a thermo-chemical model which reproduces the SED. The model provides the 2D ($R, Z$) distributions of temperature and CO abundance. Then they introduced a CO depletion factor, which is varied from 0.001 to 50 with a step-size of 1.1 in a logarithmic scale, at each radial bin, and calculated the CO flux for grid of models to find the best fit depletion factor. The uncertainty in the CO column density is thus 10 \% in most of the disk radii. Exceptions are the innermost and outermost radii; at these radii, the C$^{18}$O line fluxes have relatively large uncertainties, and the error of the CO column density is 1 sigma. They also derived the CO column densities using the $J=2-1$ transition of $^{13}$CO, and $J=1-0$ transition of $^{13}$CO, C$^{18}$O, and C$^{17}$O, which are in reasonable agreement with the value based on C$^{18}$O ($J=2-1$). 

Overall, the radial profile of the HCO$^+$ column density is similar to that of CO.
The panels in the middle row of Figure \ref{fig:col_HCOp_comb1} show the column density ratio $N$(HCO$^+$)/$N$(CO). Since the CO column density was calculated on a finer radial grid than HCO$^+$, we interpolated the CO column density at the radii of HCO$^+$. The error bars in the middle panels reflect only the error of HCO$^+$ column density, which dominates over that of CO column density. The ratio is mostly within $10^{-5}-10^{-4}$ at $\sim 100-500$ au, which suggests a close chemical link between these species, as expected from chemical models \citep[e.g.][]{aikawa15,teague15}.
%One exception is MWC~480, in which the column density ratio is less than $10^{-5}$ inside $\sim 200$ au.
%The vertical dashed line indicates the CO snowline, i.e.\ the radius at which the abundances of gaseous CO and icy CO become equal in the midplane in the thermo-chemical model of \citet{zhang20}.

In the disk of IM~Lup, the column densities of both HCO$^+$ and CO increase inwards from $R\sim 500$ au to 100 au, where the HCO$^+$/CO column density ratio is remarkably constant. The HCO$^+$ column density then declines at $\lesssim 100$ au, while the CO column density becomes flat and rises inwards at $\lesssim 50$ au.
In the GM~Aur disk, the HCO$^+$/CO column density ratio exceeds $10^{-4}$ outside the dust continuum edge. Both the HCO$^+$ and CO column densities increase inwards up to $\sim 100$ au, while they have small local humps.  At $R \sim 60$ au, the CO column density has a local maximum, while that of HCO$^+$ has a local minimum. This coincidence needs to be taken with caution, however, since the beam size of HCO$^+$ data is two times larger than that of CO, and since the column density estimates at the innermost radii suffers line broadening due to the velocity gradient within a beam.
%The CO column density then decreases inwards from $\sim 60$ au to $\sim 40$ au. Inside the expected CO snow line, the CO column density is rather uncertain. The HCO$^+$ column density is not well constrained at the innermost radial bin because the azimuthally averaged spectrum is significantly broadened by the velocity gradient within the 0\farcs3 beam.}
In the disk of AS~209, the radial profile of HCO$^+$ significantly differs from that of CO at $R\lesssim 100$ au. While the CO column density shows a broad depression at $R\sim 45-120$ au, the HCO$^+$ column density increases inwards up to $\sim 75$ au, and stays constant at inner radii within the error bars.
%Further inwards, the HCO$^+$ column density is not well constrained, while the CO column density rises sharply.
In the disk around HD~163296, the column density ratio of HCO$^+$/CO decreases inwards up to  $R\sim 240$ au, which corresponds to the edge of the millimeter dust continuum. From 240 au to 65 au, the ratio stays constant or has a shallow rise, while both the CO and HCO$^+$ column densities increase inwards. Inside the radius of $\sim 65$ au, which coincides with the CO snow line in the thermo-chemical model, the HCO$^+$ column density is roughly constant, while the CO column density increases towards the center by more than an order of magnitude. The sharp rise of CO column density and relatively flat distribution of HCO$^+$ column density inside the CO snow line is also seen in MWC~480, while the overall column density ratio of HCO$^+$/CO is slightly lower than that in HD~163296. We note that the X-ray spectrum of MWC~480 is significantly softer than that of HD~163296 \citep{dionatos19}. Since X-rays are the major ionization source in the molecular layer (see \S \ref{subsec:model}), the relatively low HCO$^+$/CO column density ratio in MWC~480 could be due to the lack of high energy X-rays ($\gtrsim$ a few keV).
%, but we have only upper limits of the HCO$^+$ column density at many radii, and the $N$(HCO$^+$)/$N$(CO) ratio is slightly less than $10^{-5}$ even at radii where the HCO$^+$ column density could be derived.

\setcounter{figure}{6}
\begin{figure}
\epsscale{1.0}
\plotone{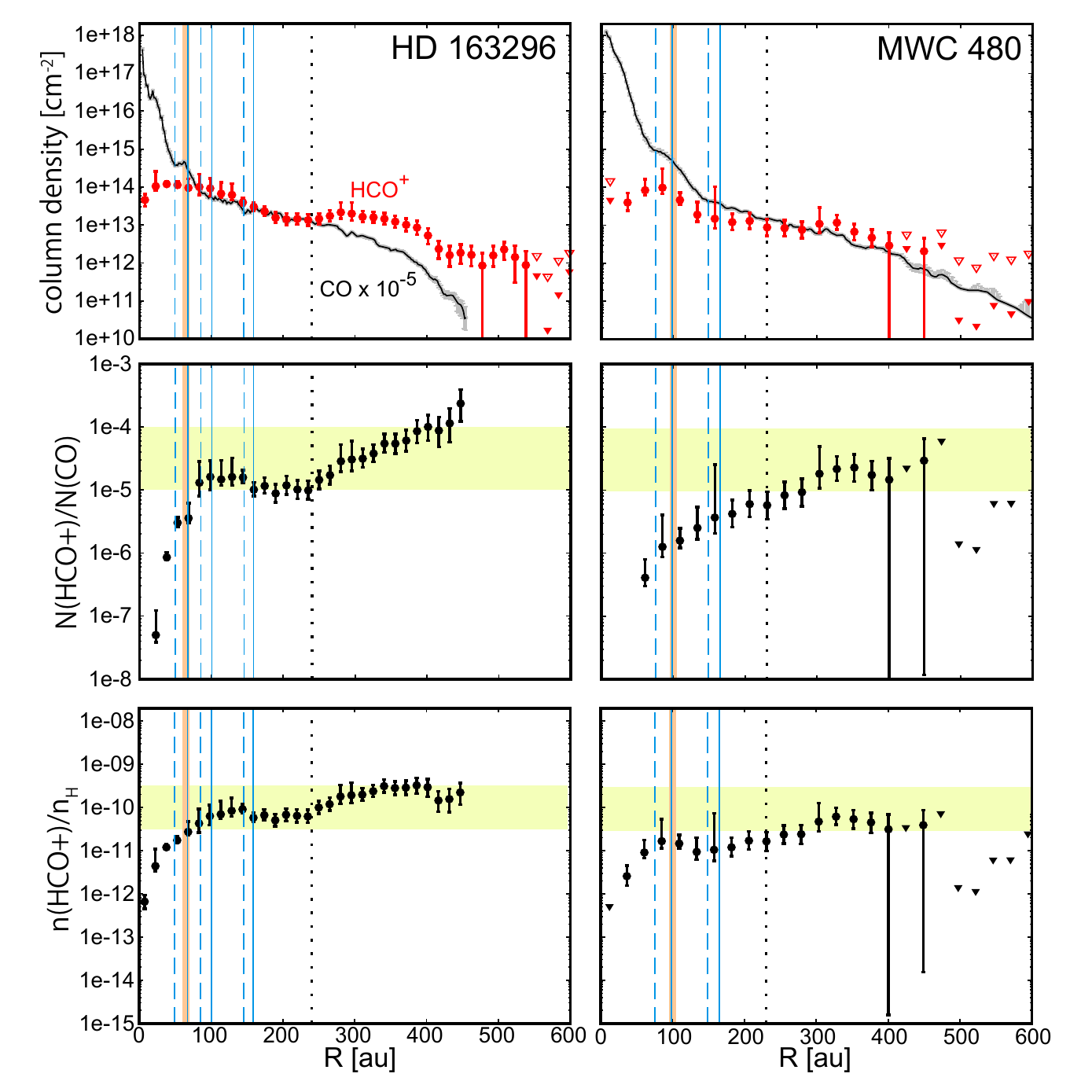}
\caption{(cont)
%\label{fig:col_HCOp_comb2}
}
\end{figure}

%\begin{figure}
%\epsscale{1.2}
%\plotone{col_HCO+_rev.pdf}
%\caption{Column densities of HCO$^+$ (red), N$_2$H$^+$ (blue), and N$_2$D$^+$ (cyan). Gray points depict the CO column density {\bf scaled by $10^{-5}$,} to be compared with HCO$^+$ column density. The error bars corresponds to {\bf 20 \% or 1 $\sigma$ for CO (see text),} while they show the 16th and 84th percentile for other species. At radii where the median value of the molecular column density is lower than the value at 84th percentile by a factor of $>10$, we plot the upper limits as inverted triangles (84 percentile for closed triangles and 98 percentile for open triangles). The vertical dashed line marks the CO snowline as inferred from the model by \citet{zhang20}.
%{\bf The caption should also say what the error bars represent.}
%\label{fig:column_HCOp}}
%\end{figure}

%\begin{figure}
%\epsscale{1.2}
%\plotone{ratio.pdf}
%\caption{The ratio of the column densities of HCO$^+$ to CO. We consider only the error in HCO$^+$ column density, since it dominates over the error of CO column density estimates in most of the disk radii.
%{\bf The caption should also say what the error bars represent.}
%\label{fig:ratio}}
%\end{figure}

\section{Discussion}\label{sec:discussion}

\subsection{{\rm HCO$^+$} abundance in the warm molecular layers}
\begin{figure}
\epsscale{1.2}
\plotone{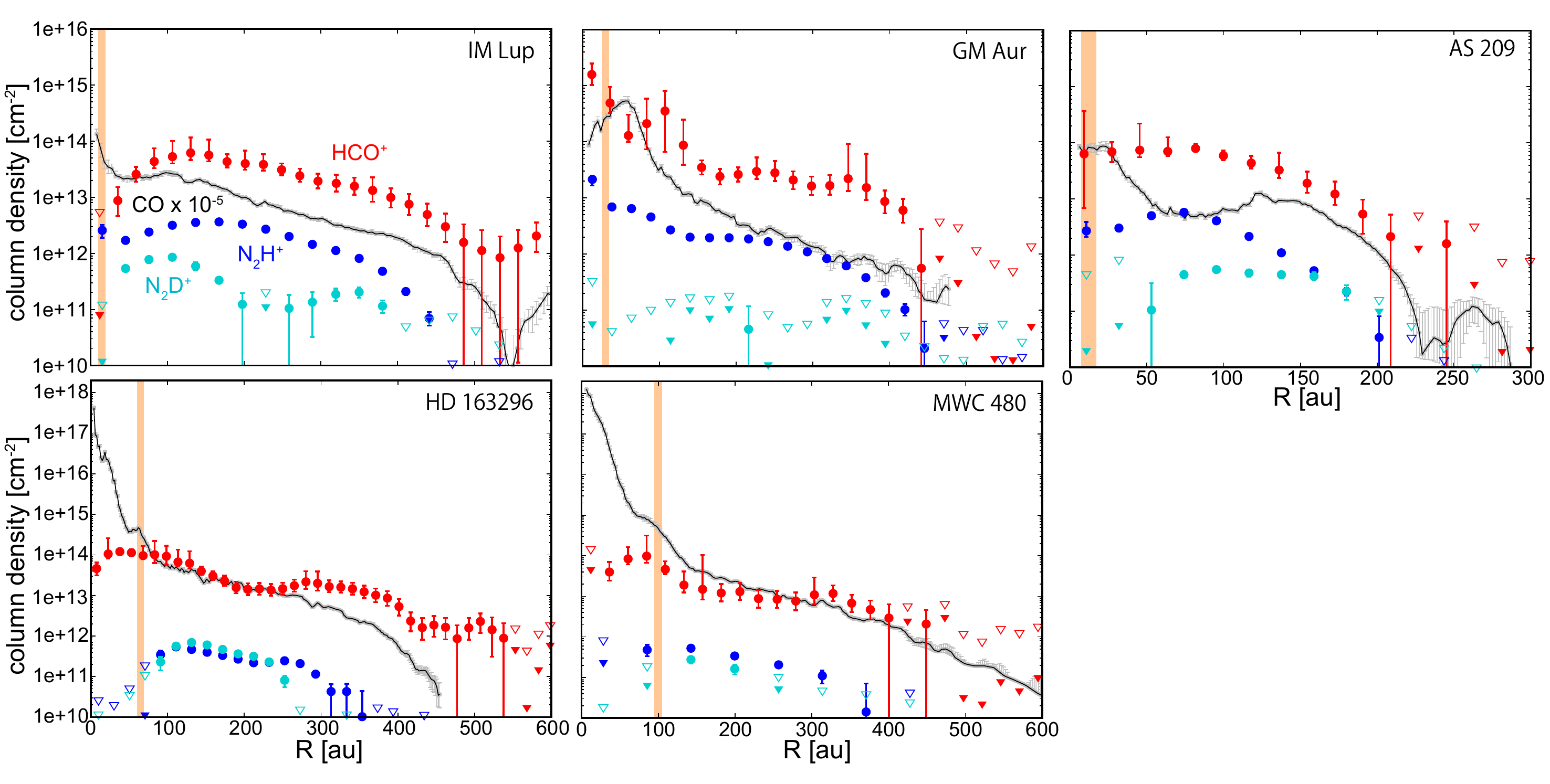}
\caption{Column densities of HCO$^+$ (red), N$_2$H$^+$ (blue), and N$_2$D$^+$ (light blue). Black lines with gray error bars depict the CO column density scaled by $10^{-5}$. The error bars corresponds to 10 \% or 1 $\sigma$ for CO (see \S 3.2), while they show the 16th and 84th percentile for other species. At radii where the median value of the molecular column density is lower than the value of the 16th percentile by a factor of $>10$, we plot the upper limits as inverted triangles (84th percentile for closed triangles and 98th percentile for open triangles). The vertical orange bars mark the CO snowline as inferred from the model by \citet{zhang20}.
\label{fig:column_HCOp}}
\end{figure}

In the molecular layers of protoplanetary disks, the major molecular ions are H$_3^+$, HCO$^+$ and N$_2$H$^+$, among which H$_3^+$ cannot be observed at millimeter wavelengths. Its deuterated counterpart o-H$_2$D$^+$ has not been detected so far \citep[e.g.][]{chapillon11}. Figure \ref{fig:column_HCOp} shows the radial profiles of the HCO$^+$, N$_2$H$^+$, and N$_2$D$^+$ column densities obtained in this work and by \citet{cataldi20}. 
%While N$_2$D$^+$ is also detected in \citet{cataldi20}, the column density ratio of N$_2$D$^+$/N$_2$H$^+$ is $\lesssim 1$.
We can see that HCO$^+$ has the largest column densities among the observable molecular ions in our five disks.
 
HCO$^+$ is expected to be the major molecular ion in the warm molecular layer where CO gas is abundant. %For example, \citet{aikawa15} showed that HCO$^+$ is more abundant than H$_3^+$ when the abundance ratio of CO to electrons is higher than $1.5\times 10^2 (T/300 {\rm K})^{-0.69}$. 
Thus, here we assume that the majority of HCO$^+$ coexists with CO in the warm molecular layer, and derive the HCO$^+$ abundance there, based on the HCO$^+$ to CO column density ratio. Furthermore, we assume that the CO abundance relative to hydrogen nuclei is equal to its canonical abundance in molecular clouds, $5.0\times 10^{-5}$, multiplied by the CO depletion factor derived in \cite{zhang20}. The bottom panels in Figure \ref{fig:col_HCOp_comb1} show the estimated HCO$^+$ abundance.
Since the uncertainty (error) of the HCO$^+$ column density dominates over the that of the CO column density at most disk radii, and since the methods of error estimation are different between HCO$^+$ and CO, we evaluate the error of the HCO$^+$ abundance considering only that of the HCO$^+$ column density.
%The error of the HCO$^+$ abundance is calculated by error propagation.
Overall, the HCO$^+$ abundance is in a range of $3\times 10^{-11}-3\times 10^{-10}$
%, both at different radii within individual disks, and 
outside $\sim 100$ au, and tends to decline towards the disk center at the inner radii.

In those inner regions, the CO-rich molecular layer extends towards the midplane, although detailed analysis by \citet{zhang20} shows that the gaseous CO fractional abundance does not necessarily recover its canonical value (i.e.\ $10^{-4}$ relative to H$_2$) even inside of the CO snow line. 
Closer to the midplane, the gas density is higher and the ionization degree, i.e.\ the HCO$^+$ abundance, should be lower. Furthermore, HCO$^+$ is suppressed by grain-surface recombination (i.e.\ collision with a negatively charged grain), which could be more efficient than the recombination in the gas phase at high densities. The HCO$^+$ abundance averaged over the CO gas rich layer should thus be lower around and inside the CO snow line than at outer radii (see \S \ref{subsec:model}).

Although HCO$^+$ is expected to be one of the major molecular ions in the warm molecular layer, it is not always the most abundant ion. The relative abundances of major molecular ions, H$_3^+$, HCO$^+$, and N$_2$H$^+$, are determined by the balance between proton transfer and recombination of relevant species, i.e.\ H$_2$, CO, and N$_2$ \citep{aikawa15}. Furthermore, if the abundance of small grains is low enough to allow significant penetration of UV radiation, atomic ions such as C$^+$ and S$^+$ can dominate even in the CO rich layers \citep[e.g.][]{aikawa15}. Therefore, the HCO$^+$ abundance derived here should be considered as a lower limit of the ionization degree in the molecular layer.

\subsection{ \texorpdfstring{HCO$^+$}{HCO+} enhancement in gas gaps}
%KOKOKARA
The HCO$^+$ abundance exceeds $10^{-9}$ at 50 au $\lesssim R \lesssim 150$ au in AS~209, where the CO column density shows a depression. %Figure \ref{fig:col_HCOp_comb1} shows that the column density of HCO$^+$ has a local maximum around $R\sim 80$ au, where the CO column density has a local minimum. 
%Although the CO depletion factor also has its local minimum around this radius, the radial variation of $N$(HCO$^+$)/$N$(CO) is more significant than that of CO depletion factor, which results in the relatively high HCO$^+$ abundance. 
\citet{favre19} observed DCO$^+$ $J=3-2$ in AS 209 with high resolution (0\farcs26 $\times$ 0\farcs21) and found a ring-like emission with its peak at $\sim 85$ au. They suggest that the over-density of DCO$^+$ is caused by the more efficient ionization at the radius of gas and dust depletion, which is carved by a planet of $\lesssim 0.3$ M$_{\rm Jup}$ at $\sim 100$ au. The high HCO$^+$ abundance at 50 au $\lesssim R \lesssim 150$ au is consistent with their scenario.

\citet{alarcon20}, on the other hand, calculated the thermal structure in two disk models for AS~209; model A assumes a smooth gas distribution and a CO abundance drop around $R\sim 80$ au, while model B assumes a constant CO abundance combined with a gas density drop around 80\,au. They found that the radial pressure profile derived by \citet{teague18b} is shallower than in model B, and concluded that the CO column density drop at 80\,au is caused mainly by a CO abundance drop rather than a gas density drop. Self-consistent modeling of the thermal structure
and the HCO$^+$ and DCO$^+$ chemistry would be useful to distinguish between these scenarios. 
%In such a modeling, the abundance and distribution of small dust grains could also be a key parameter, as it affects both the gas temperature and the photoionization. The latter produces electrons, which destroy HCO$^+$.
%; if HCO$^+$ is the dominant charge carrier, its abundance would be higher at lower gas density, since the ionization degree is proportional to $n_{\rm H}^{-1/2}$.

We also note correlated local enhancements of HCO$^+$ and DCO$^+$ in HD 163296. \citet{flaherty17} observed DCO$^+$ ($J=3-2$) with a beam of $0.5'' \times 0.59''$ and found that DCO$^+$ emission is confined to three concentric rings at 54, 124, and 214 au. The ring at 124 au is the brightest and coincides with the shoulder seen in HCO$^+$ emission in Figure \ref{fig:radial_profile} and the local enhancement of the HCO$^+$/CO column density ratio (Figure \ref{fig:col_HCOp_comb1}).
\citet{teague18a} analyzed the disk gas kinematics around HD~163296 and found deviations from Keplerian rotation, which suggests the presence of a 1 $M_{\rm Jup}$ planet at 83 au and a 1.3 $M_{\rm Jup}$ planet at 137 au. Hydrodynamic models of the HD~163296 disk with these planets show gas gaps (i.e.\ gas density decrease) around the planet orbits.
 The local enhancement of DCO$^+$ and HCO$^+$ around 125 au could be related to such gas gaps.
 
In MWC 480, the CO column density shows a local minimum around the gap seen in dust continuum at $\sim 76$ au (Figure \ref{fig:col_HCOp_comb1}). We note that the HCO$^+$ column density has a local maximum at this radius; we derived the column density distribution with radial grids of a quarter of the beam size to confirm the coincidence.
It again suggests a correlation between gas gap and HCO$^+$ enhancement. Observations of HCO$^+$ with higher angular resolution are desirable for further studies.
 
%\begin{figure}
%\epsscale{1.2}
%\plotone{abun_HCO+_rev.pdf}
%\caption{Abundance of HCO$^+$ relative to hydrogen nuclei in the CO gas rich layer. The vertical dashed lines depict the radius of the CO snow line obtained via thermo-chemical modeling by \citet{zhang20}. The horizontal dashed-dotted lines depict the abundance of $1\times 10^{-10}$. At radii where the median value of the HCO$^+$ column density is lower than the value at 84th percentile by a factor of $>10$, we plot the 1-$\sigma$ upper limit with an inverted triangle.
%\label{fig:abun_HCOp}}
%\end{figure}

\subsection{Constraining the ionization structure with a template disk chemistry model}\label{subsec:model}
 
MAPS observations, as well as previous ALMA observations of disks in the last few years, show that detailed physical structure, e.g.\ size, flaring, and temperature structure vary significantly among disks \citep[e.g.][]{law20_surfaces_vertical_distributions}. While source-specific modeling is desirable, it is out of the scope of the present work.
However, we note that the derived abundance and its radial distribution of HCO$^+$ have common features; the HCO$^+$ abundance is $\sim 3\times 10^{-11} - 3 \times 10^{-10}$ at $\gtrsim 100$ au, while it tends to decrease inwards at smaller radii. These features should reflect the basic chemistry of HCO$^+$, which is not very sensitive to the details of the disk structure. Therefore, we compare our results with some template disk models.

We adopt the disk chemistry model of \citet{aikawa18}; the disk mass is $1.7\times 10^{-2} M_{\odot}$ and the mass of the central star is 0.5 $M_{\odot}$. Stellar UV and X-ray luminosities are $10^{31}$ erg s$^{-1}$ and $10^{30}$ erg s$^{-1}$, respectively. 
%The gas-to-dust mass ratio is set to be $10^{-2}$. 
While the dust sedimentation is not considered, the maximum grain size is set to be 1 mm. Since the total gas-to-dust mass ratio is set to be the same as the interstellar value ($\sim 100$), the abundance of small dust grains is reduced compared with the interstellar dust.
The 2D ($R$,$Z$) distributions of the gas number density (i.e.\ number density of hydrogen nuclei $n_{\rm H}$) and temperature are shown in Figure \ref{fig:model_rel_abun_appendix} in Appendix \ref{appendix:model}.
While \citet{aikawa18} varied several parameters in the calculation of the chemistry model, we here show three models: their fiducial model, a high C/O model, and a low $\zeta$ model, which is newly calculated for the present work. The fiducial model is a static disk model with a cosmic-ray ionization rate of $5\times 10^{-17}$ s$^{-1}$. Two other ionization sources, X-rays and the decay of SLR are included in the model with the X-ray spectrum of TW~Hya and the ionization rate by SLR of $\zeta_{\rm SLR}=1\times 10^{-18}$ s$^{-1}$. The initial molecular abundance for the disk chemistry is set by calculating the chemical evolution from the molecular cloud formation stage to the collapse of a star-forming core. 
In the fiducial model, the elemental abundance of C and O are $7.82\times 10^{-5}$ and $1.8\times 10^{-4}$ relative to hydrogen nuclei (i.e.\ C/O=0.43). Disk observations in recent years, however, suggest low C/H ratios and high C/O ratios in the gas phase \citep[e.g.][]{bergin14,bergin16,zhang20, bosman21}. This is likely due to the removal of ice by the sedimentation and radial migration of ice-coated pebbles \citep{kama16, krijt18}. Thus, in the high C/O model, H$_2$O is completely removed and CO is reduced by one order of magnitude in the initial molecular abundances, which results in a C/O ratio of 1.43. In addition to these two models, we run a low $\zeta$ model, in which the cosmic-ray ionization rate is set to zero. Ionization rates by X-ray and SLR are set to be the same as in the fiducial model, and the initial molecular abundance is the same as the high C/O model. Since we aim to constrain the ionization rate rather than identifying the main ionization source, the low $\zeta$ model can simply be considered as a model with a midplane ionization rate of $\zeta_{\rm mid}=10^{-18}$ s$^{-1}$, while $\zeta_{\rm mid}$ is $5\times 10^{-17}$ s$^{-1}$ in the fiducial model and high C/O model.

While molecular evolution in the disk is calculated up to $t=1\times 10^6$ yr, we compare our observational results with the molecular abundances at $t=1\times 10^5$ yr, which is comparable to the vertical mixing timescale at the radius of several tens of au assuming an $\alpha$ parameter of $10^{-3}$ for turbulent mixing \citep[e.g.][]{aikawa96}.
%because radial accretion and/or mixing are not explicitly taken into account in the model.
In gas-grain chemical networks, volatile species tend to be converted to less volatile ices. At $t=10^6$ yr, a significant amount of CO and N$_2$ is converted to CO$_2$, CH$_3$OH, and NH$_3$, which could be an artifact of the static disk assumption (i.e.\ without mixing or accretion). The desorption rate and thermal diffusion rate (and thus the reaction rate) on the grain surface are very sensitive to temperature. In the static model, in which gas and dust stay at the same position, and thus are kept at a constant temperature, only a limited number of species can effectively diffuse and react, which results in an accumulation of specific products.
At $t=1\times 10^5$ yr, such accumulations are less significant \citep[see][for more details]{aikawa15}.

\begin{figure}
\epsscale{1.2}
\plotone{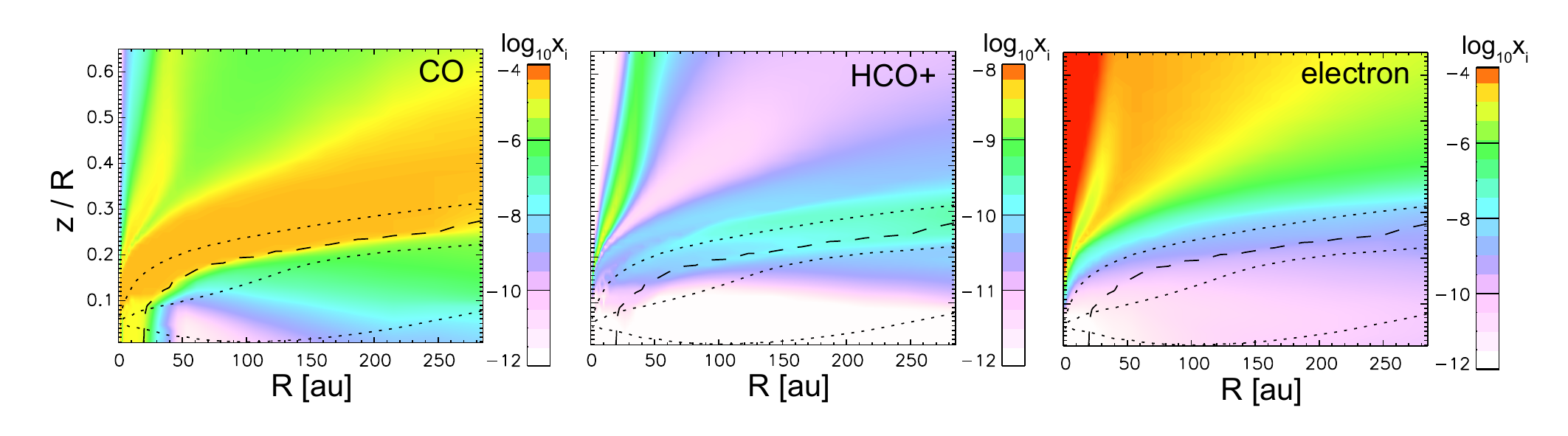}
\caption{
The abundances of CO (left), HCO$^+$ (middle), and electrons (right) relative to hydrogen nuclei in the low $\zeta$ model. The dashed line depicts the CO snow surface and the dotted lines show the positions where the X-ray ionization rate is $5\times 10^{-17}$ s$^{-1}$ (the upper dotted line), $1\times 10^{-18}$ s$^{-1}$ (the middle dotted line), and $1\times 10^{-19}$ s$^{-1}$ (the lower dotted line),
\label{fig:model_rel_abun_lowzeta}}
\end{figure}

Figure \ref{fig:model_rel_abun_lowzeta} shows the abundances of CO, HCO$^+$, and electrons relative to hydrogen nuclei, in the low $\zeta$ model. The abundances in the other models are shown in Figure \ref{fig:model_rel_abun_appendix} in Appendix \ref{appendix:model}. 
%We note that the molecular abundance is relative to the hydrogen nuclei in the models, while we evaluated the HCO$^+$ abundance relative to H$_2$ from the observational data, which causes a factor of 2 difference.
We can see that CO and HCO$^+$ are mostly co-spatial and that the observed features are reproduced; the HCO$^+$/CO abundance ratio is $\sim 10^{-5}-10^{-4}$, and the HCO$^+$ abundance is $\sim 10^{-10}$ in the lower part of the CO-rich layer, where the density is higher. In Figure \ref{fig:model_rel_abun_lowzeta}, the dashed line depicts the CO snow surface and the dotted lines depict the position where the X-ray ionization rate is $5\times 10^{-17}$ s$^{-1}$ (upper), $1\times 10^{-18}$ s$^{-1}$ (middle), and $1\times 10^{-19}$ s$^{-1}$ (lower). We note that the dashed line and dotted lines are roughly parallel at radii $>50$ au, probably because both parameters are controlled by the radiation transfer from the central star (i.e.\ stellar irradiation and X-rays). Inside the CO snow line (where the dashed line crosses the midplane), CO is indeed abundant near the midplane, although its abundance is reduced by an order of magnitude compared to the initial value ($\sim 10^{-5}$) due to the conversion to less volatile species such as CO$_2$ ice. In this region, the HCO$^+$/CO abundance ratio is less than $10^{-5}$ in the low $\zeta$ model, which is also consistent with the observations. Comparing the HCO$^+$ and the electron abundances, we can see that the former is slightly lower than the latter even in the layers where the HCO$^+$ abundance peaks. This means that the ionization degree is actually higher than the HCO$^+$ abundance; atomic ions are relatively abundant even in the CO-rich layer in the model.

\begin{figure}
\epsscale{1.2}
\plotone{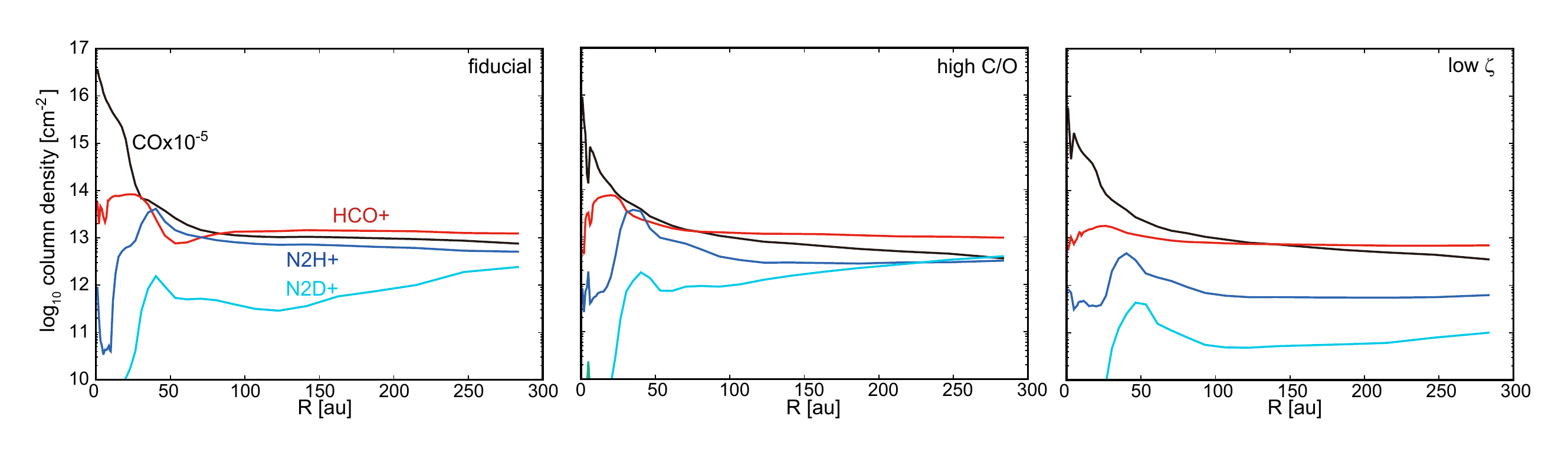}
\caption{
Column densities of CO (multiplied by $10^{-5}$), HCO$^+$, N$_2$H$^+$, and N$_2$D$^+$ in the fiducial model (left), high C/O model (middle) and low $\zeta$ model (right).
\label{fig:model_col}}
\end{figure}

\begin{figure}
\epsscale{1.2}
\plotone{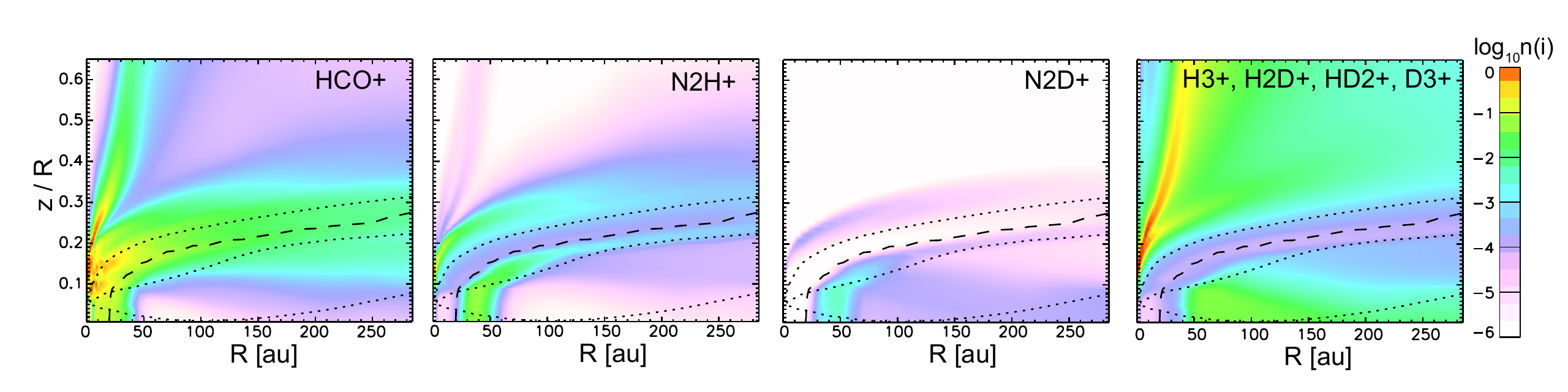}
\caption{
The absolute abundances (number densities) of HCO$^+$, N$_2$H$^+$, N$_2$D$^+$, and H$_3^+$ (and its deuterated isotopomers) relative to hydrogen nuclei in the low $\zeta$ model. The dashed line depicts the CO snow surface and the dotted lines depict the position where the X-ray ionization rate is $5\times 10^{-17}$ s$^{-1}$ (the upper dotted line), $1\times 10^{-18}$ s$^{-1}$ (the middle dotted line), and $1\times 10^{-19}$ s$^{-1}$ (the lower dotted line),\label{fig:model_abs_ion_lowzeta}}
\end{figure}

Figure \ref{fig:model_col} shows the radial profiles of the column densities of CO (multiplied by $10^{-5}$), HCO$^+$, N$_2$H$^+$, and N$_2$D$^+$ in the three models.
As references, 2D distributions of absolute abundances (i.e.\ number densities) of major molecular ions in the low $\zeta$ model are shown in Figure \ref{fig:model_abs_ion_lowzeta}. Similar plots for the other two disk models are shown in Figure \ref{fig:model_abs_ion_appendix} in Appendix \ref{appendix:model}.
The column densities in the models are almost constant outside $\sim 100$ au, while the observed values decline outwards. This discrepancy might arise because the model gas density is not exponentially tapered, but simply follows a power law up to a radius of 284 au.
We thus only compare the model column densities inside $\sim 200$ au with the observed values.
In all three models, the radial profile of the column density ratio $N$(HCO$^+$)/$N$(CO) is similar to those in Figure \ref{fig:col_HCOp_comb1}; $N$(HCO$^+$)/$N$(CO) is nearly constant ($\sim 10^{-5}$), but declines towards the disk center inside a certain radius, which corresponds to the CO snow line in the disk models.

%In the disks around T Tauri stars, $N$(HCO$^+$)/$N$(CO) in the innermost radii is larger than that in the two Herbig Ae disks, and is more consistent with the models with high C/O ratio (i.e. the high C/O model and low $zeta$ model) than with the fiducial model.

%Finally, we compare the column densities of molecular ions in template models those derived in the present work (HCO$^+$) and \citet{cataldi20} (N$_2$H$^+$ and N$_2$D$^+$).
Figure \ref{fig:model_col} shows that at radii outside the CO snow line ($\sim 20 $ au in the template model) the HCO$^+$ column density is not sensitive to $\zeta_{\rm mid}$, since the contribution of the X-ray ionized region is significant in the CO-rich layer there (see Figure \ref{fig:model_abs_ion_lowzeta} and Figure \ref{fig:model_abs_ion_appendix}).
Inside the CO snow line, the CO-rich layer extends to the midplane. The HCO$^+$ column density and the gas column density ratio $N$(HCO$^+$)/$N$(CO) around and inside the CO snow line is thus more sensitive to $\zeta_{\rm mid}$.
\footnote{We note that the absolute abundance of HCO$^+$ in the midplane is high even at radii slightly outside the CO snow line. For HCO$^+$ to be the abundant ion, CO does not need to fully sublimate to the gas phase \citep{aikawa15}.}
The peak value of the HCO$^+$ column density is lower in the low $\zeta$ model than in the models with $\zeta_{\rm mid}=5\times 10^{-17}$ s$^{-1}$ by a factor of 4.6 (fiducial model) and 4.2 (high C/O model). The column density ratio of HCO$^+$/CO at the HCO$^+$ column density peak is also lower in the low $\zeta$ model accordingly.
Around the CO snow lines predicted from the thermo-chemical model of \cite{zhang20}, the HCO$^+$ column density derived from the observation is $\gtrsim 10^{14}$ cm$^{-2}$ with the column density ratio of HCO$^+$/CO $\sim 10^{-5}$ in our targeted disks, as in our fiducial model or high C/O model, except for those around IM Lup and MWC 480. It may  suggest $\zeta_{\rm mid} \sim 10^{-17}$ s$^{-1}$ around the CO snow line.
We note, however, that the HCO$^+$ column density would also depend on the total gas column density, which is not varied in the set of template models shown here. While the mass of the template disk model is within the range of estimated values for GM~Aur and IM~Lup, AS~209 is less massive and the disks around the Herbig Ae stars are more massive \citep{zhang20} (see \S 4.4). Source-specific models are needed for further comparison.
%But if the gas column density (i.e.\ disk gas mass) is higher, the CO column density should also be higher. The dependence on the gas column density should be canceled (at least partially) by considering the HCO$^+$/CO ratio.
%Assuming that the gas density in the CO-rich layer is proportional to the gas column density, $N$(HCO$^+$)/$N$(CO) would be inversely proportional to the square root of the gas column density, since the ionization rate is proportional to $n_{\rm H}^{-1/2}$.
%In the observations, $N$(HCO$^+$)/$N$(CO) at the radius of the HCO$^+$ column density peak is higher than $10^{-5}$, which is more consistent with the template models with $\zeta_{\rm mid}=5\times 10^{-17}$ s$^{-1}$ than the low $\zeta$ model. If the $\zeta_{\rm mid}$ is smaller, the gas density near the midplane (and thus the gas column density) should be lower accordingly.
%One exception is MWC~480; the lower $N$(HCO$^+$)/$N$(CO) than other 4 disks indicates lower $\zeta_{\rm mid}$.
%We note that this conclusion is not sensitive to the CO abundance in the warm midplane, as long as it is abundant enough to make HCO$^+$ the most abundant molecular ion.

\subsection{Ionization rate traced by \texorpdfstring{N$_2$H$^+$}{N2H+} and \texorpdfstring{N$_2$D$^+$}{N2D+}}

The template disk models (Figures \ref{fig:model_col},  \ref{fig:model_abs_ion_lowzeta}, and \ref{fig:model_abs_ion_appendix}) indicate that, in contrast to HCO$^+$, the column densities of N$_2$H$^+$ and N$_2$D$^+$ are more sensitive to the ionization rate below the CO snow surface, where these molecular ions are relatively abundant. In the cold midplane, H$_3^+$ and its deuterated isotopomers are the most abundant ions, but H$_3^+$ and D$_3^+$ are not observable at radio wavelengths, and H$_2$D$^+$ and HD$_2^+$ are not detected so far, possibly due to unfavorable o/p ratios \citep[e.g.][]{chapillon11}. Figure \ref{fig:model_abs_ion_appendix} shows that among the observable molecular ions, N$_2$D$^+$ is the best proxy of deuterated H$_3^+$ and thus the best probe of ionization rate in the midplane. Although its abundance is lower than the sum of H$_3^+$ and its deuterated isotopomers, the gas above the CO snow surface does not significantly contribute to its column density in our template disk models. In the template disk models, the N$_2$H$^+$ and N$_2$D$^+$ column densities have a peak around $R\sim 50$ au, where these molecular ions are abundant in the midplane \citep[see also][]{qi19}. This indicates that $\zeta_{\rm mid}$ could be best investigated by their peak column densities, especially by the N$_2$D$^+$ column density.

The column densities of N$_2$H$^+$ and N$_2$D$^+$ in our target disks are derived by \citet{cataldi20} and are plotted in Figure \ref{fig:column_HCOp}. As expected from the template disk models, these column densities have a peak outside the CO snow line, except for GM Aur, for which the N$_2$H$^+$ column density is the highest at the innermost radial grid. The value at the innermost grid could, however, be overestimated, since these column densities are derived by assuming a fixed excitation temperature of 20\,K. When the excitation temperature is set to be the same as the midplane temperature of the thermochemical model of \citet{zhang20}, the N$_2$H$^+$ column density has a peak ($\sim 10^{13}$ cm$^{-2}$) around $R\sim 100$ au \citep{cataldi20}.

\begin{figure*}
\epsscale{0.7}
\plotone{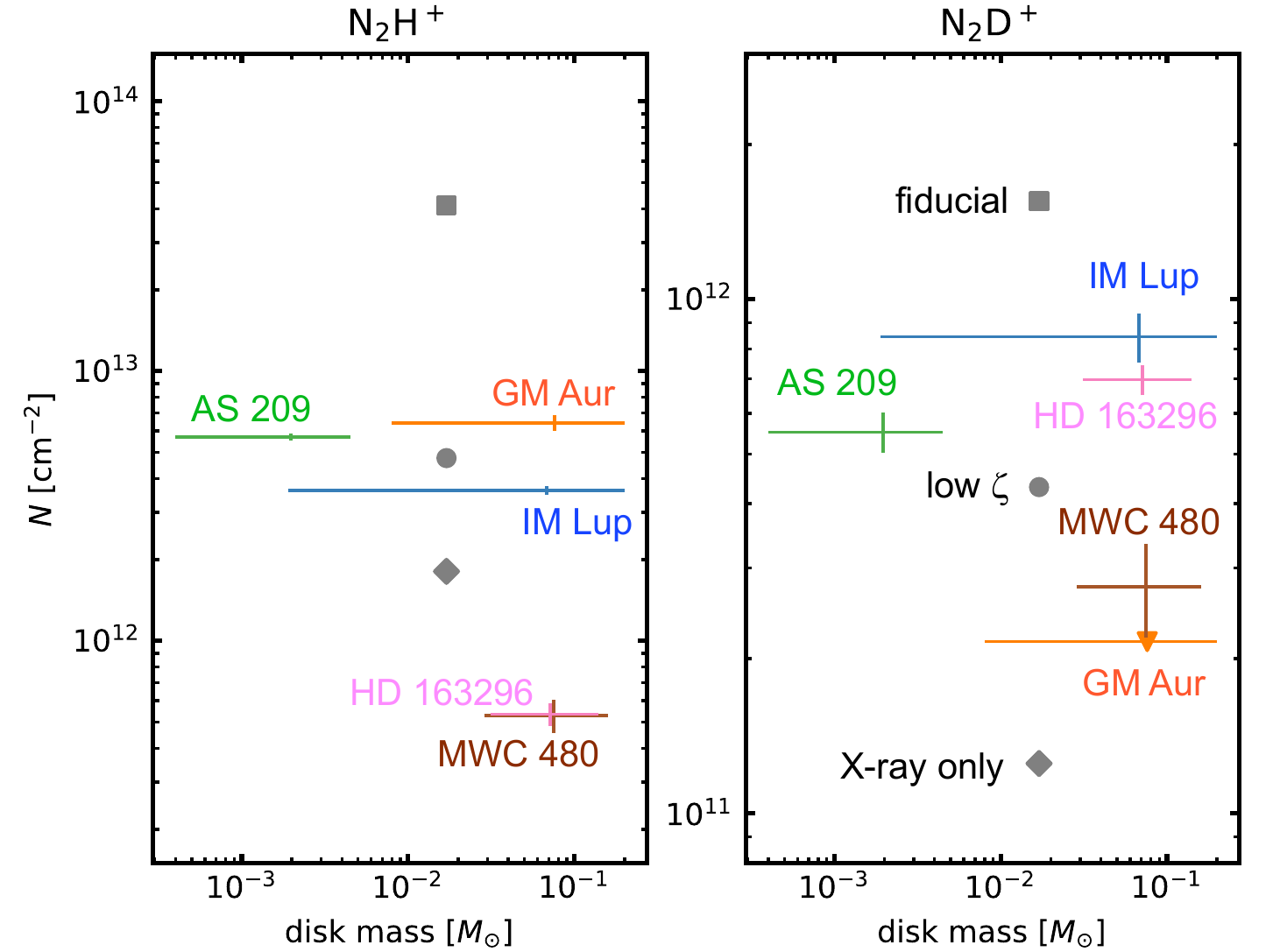}
\caption{The \nthp and \ntdp column densities as a function of disk mass. For each disk, a mass range adopted from \citet{zhang20} is shown as horizontal error bar. The column densities correspond to the peak values, with the error bars indicating the 16th and 84th percentiles. For \ntdp in GM~Aur, we instead plot the 99.85th percentile at $\sim$100\,au as an upper limit. Predictions from the template models are shown by the gray points. \label{fig:N2H+_N2D+_vs_diskmass}}
\end{figure*}

Unlike HCO$^+$, the distributions of N$_2$H$^+$ and N$_2$D$^+$ are expected to anti-correlate with that of CO. We thus consider their column densities rather than the column density ratio to CO.
In Figure \ref{fig:N2H+_N2D+_vs_diskmass}, we plot the peak \nthp and \ntdp column densities derived by \citet{cataldi20} versus disk mass to see if there is any trend. For each disk, a range of possible masses is shown, extending from the minimum masses to the best-fit mass given in Tables 3 and 2 of \citet{zhang20}, respectively.
While the estimated disk masses of IM Lup, GM Aur, HD 163296, and MWC 480 overlap with each other\footnote{\citet{fedele18} reproduced their 1.3 mm dust continuum image with a disk model with a dust mass of $3.5\times 10^{-4}$ M$_{\odot}$. If the gas/dust mass ratio is 100, the disk mass of AS 209 is also similar to the disk mass of IM Lup and GM Aur.}, their N$_2$H$^+$ and N$_2$D$^+$ peak column densities vary, which indicates that the midplane ionization rate varies among the disks.
The gray squares and circles depict the peak column densities around $R\sim 50$ au in the fiducial model and low $\zeta$ model, respectively. The gray diamonds show the values in the model with X-ray ionization only; i.e.\ the midplane ionization rate is $\lesssim 10^{-19}$ s$^{-1}$. The peak column densities of the high C/O model are not plotted but are similar to those of the fiducial model.

The N$_2$H$^+$ column densities of IM~Lup, GM~Aur, and AS~209 indicate $\zeta_{\rm mid}\sim 10^{-18}$ s$^{-1}$. For MWC 480 and HD 163296, the N$_2$H$^+$ column density is lower than that of the X-ray only model, which indicates that the X-ray ionization is less effective in these disks than in our template models. We note that a fraction of N$_2$H$^+$ exists in the layer with relatively high X-ray ionization rate ($\gtrsim 5\times 10^{-17}$ s$^{-1}$), which sets the floor value of the N$_2$H$^+$ column density in our template models. For MWC 480, the low N$_2$H$^+$ column density is consistent with its soft X-ray spectrum and relatively low HCO$^+$ column density. Alternative explanation for the variety of N$_2$H$^+$ column density would be the temperature structure in the disk. The column density of N$_2$H$^+$ should be lower if the midplane layer of the low temperature ($\lesssim 20$ K) is thinner \citep{qi19}. This could be the case for MWC 480 disk, in which both N$_2$H$^+$ and N$_2$D$^+$ column densities are low. HD 163296, on the other hand, has relatively high column density of N$_2$D$^+$, which indicate that there are plenty of cold gas in the midplane (see below).

The N$_2$D$^+$ peak column densities of IM~Lup, AS~209, and HD~163296 are similar to or slightly larger than the value from the low $\zeta$ model, which suggests $\zeta_{\rm mid}\gtrsim 10^{-18}$ s$^{-1}$.
In the HD 163296 disk, the peak N$_2$D$^+$ column density is relatively high, while the low N$_2$H$^+$ column density indicates lower X-ray ionization rate than the template models. The midplane ionization rate, which is better traced by N$_2$D$^+$, would thus be set by SLR or high energy particles. In GM Aur, on the other hand, the upper limit of N$_2$D$^+$ column density indicates a low ionization rate ($< 10^{-18}$ s$^{-1}$) in the midplane.

Finally we note that the N$_2$H$^+$ and N$_2$D$^+$ column densities depend  not only on $\zeta_{\rm mid}$, but also on other parameters such as the vertical temperature distribution and the disk mass \citep{cleeves14, qi19}. Source specific models are needed to estimate the midplane ionization rate more quantitatively for each object.

%\begin{figure}
%\plotone{column_1e5.pdf}
%\caption{Column densities of CO, HCO$^+$, and N$_2$H$^+$ in template disk models by \citet{aikawa18}. The left panel shows the fiducial model that includes ionisation by X-rays, CR and SLR. The middle panel shows a model without CR. The right panel shows a model with CO depletion, resulting in a C/O ratio of 1.43.
%\label{fig:model}}
%\end{figure}

%\section{Gas in gaps}

\section{Conclusions}\label{sec:conclusions}

We observed and analyzed HCO$^+$ $J=1-0$ and H$^{13}$CO$^+$ $J=1-0$ lines towards the protoplanetary disks around IM~Lup, GM~Aur, AS~209, HD~163296, and MWC~480. H$^{13}$CO$^+$ $J=1-0$ was detected in all five disks,
while H$^{13}$CO$^+$ $J=1-0$ was detected (SNR$>6 \sigma$) towards GM~Aur and HD 163296 and tentatively detected (SNR$>3 \sigma$) towards the other disks by a matched filter analysis in the $uv$ plane.
%not detected in the integrated intensity map, radial emission profile, or azimuthally averaged spectra. The H$^{13}$CO$^+$ line is, however, 
The disk-integrated flux of H$^{13}$CO$^+$ $J=1-0$ is also above 3$\sigma$ in IM~Lup, GM~Aur, and HD~163296.

We derived the HCO$^+$ column density by fitting the azimuthally averaged spectra of HCO$^+$ and H$^{13}$CO$^+$ simultaneously for radial bins of half the beam FWHM. %Although the H$^{13}$CO$^+$ emission is not detected in these spectra, it is useful to constrain the HCO$^+$ column density.
In all five disks, the HCO$^+$ column density increases inwards, but becomes flat or drops towards the center inside a radius of $R\sim 100$ au.
The column density ratio of $N$(HCO$^+$)/$N$(CO) is about $10^{-5}-10^{-4}$ at $R\gtrsim 100$ au, except for the MWC~480 disk, in which the column density ratio is $< 10^{-5}$ at $R\lesssim 200$ au.

We derived the HCO$^+$ abundance in the warm CO-rich layer, where HCO$^+$ is expected to be the dominant molecular ion, via the column density ratio $N$(HCO$^+$)/$N$(CO) using the radial profiles of CO column density and CO depletion factor from \citet{zhang20}. Beyond $\sim 100$ au, the derived HCO$^+$ abundance ranges from $3\times 10^{-11} - 3 \times 10^{-10}$ in the IM Lup, GM Aur, HD 163296 disks. The HCO$^+$ abundance is lower in MWC 480, possibly due to the lack of high energy ($>$ a few keV) X-rays. The HCO$^+$ abundance tends to decline at towards the disk center for $R\lesssim 100$ au. This can be explained by the lower ionization degree in denser gas, especially inside the CO snow line, where the CO-rich layer is in the midplane.

We find a hint of a correlation between the HCO$^+$ abundance and the gap carved by a putative planet in AS~209: the HCO$^+$ abundance exceeds $10^{-9}$ at $R\sim 50-150$ au, where the CO column density is depressed. This region seems to correspond to the gas gap, where ionization would be more efficient.
In HD~163296, the HCO$^+$ abundance shows a shallow bump at $100-150$ au, which coincides with the radius of DCO$^+$ enhancement found by \citet{flaherty17}. This feature could also be related to the enhanced ionization around a gas gap. In MWC 480, the HCO$^+$ column density has a local maximum at $\sim 76$ au, which coincides with the local depression of CO and the gap seen in the dust continuum.

Finally, we compared the column densities of HCO$^+$, N$_2$H$^+$, and N$_2$D$^+$ with those of template disk models: a fiducial disk model with a midplane ionization rate $\zeta_{\rm mid}$ of $5\times 10^{-17}$ s$^{-1}$, a model with CO and H$_2$O depletion, and a model with $\zeta_{\rm mid}=1\times 10^{-18}$ s$^{-1}$.
The almost constant HCO$^+$ abundance at $R\gtrsim 100$ au is explained by X-ray ionization in the CO-rich layer. The decline of the HCO$^+$ abundance at the inner radii, on the other hand, can be explained by CO sublimation inside the CO snow line; the CO-rich layer then extends to the midplane, where the ionization degree is low due to high density. %The HCO$^+$ column density derived from the observational data around the CO snow line may indicate $\zeta_{\rm mid}\sim 10^{-17}$ s$^{-1}$ in the disk midplane.
While the estimated disk mass range of IM~Lup, GM~Aur, HD~163296, and MWC~480 overlap with each other, their peak column densities of N$_2$H$^+$ and N$_2$D$^+$ vary, which may indicate that the midplane ionization rate varies among disks.
The peak N$_2$D$^+$ column density suggests a midplane ionization rate of $\gtrsim 10^{-18}$ s$^{-1}$ for IM~Lup, AS~209, and HD~163296, while the upper limit of N$_2$D$^+$ column density indicates it is $< 10^{-18}$ s$^{-1}$ for GM Aur. The peak column density of N$_2$H$^+$ is lower in MWC~480 than in other disks, which is consistent with its soft X-ray spectrum and relatively low HCO$^+$ column density. Alternatively, the low column densities of N$_2$H$^+$ and N$_2$D$^+$ in the MWC 480 disk could be due to its warmer temperature; i.e. cold miplane layer is thinner than in other disks.
Source-specific models are needed for further evaluation of the midplane ionization rate.
%while the constraint is weaker for the disks around Herbig Ae stars HD~163296 and MWC~480. We see no correlation between disk mass and the N$_2$H$^+$ and N$_2$D$^+$ column densities, 

\acknowledgments
This paper makes use of the following ALMA data: ADS/JAO.ALMA\#2018.1.01055.L,\\ADS/JAO.ALMA\#2015.1.00678.S,\\ ADS/JAO.ALMA\#2012.1.00681.S,\\ADS/JAO.ALMA\#2015.1.00657.S. ALMA is a partnership of ESO (representing its member states), NSF (USA) and NINS (Japan), together with NRC (Canada), MOST and ASIAA (Taiwan), and KASI (Republic of Korea), in cooperation with the Republic of Chile. The Joint ALMA Observatory is operated by ESO, AUI/NRAO and NAOJ. This research has made use of NASA’s Astrophysics Data System and the SIMBAD database, operated at CDS, Strasbourg, France.

We would like to thank the anonymous referee for careful reading of our manuscript and for constructive comments.
Y.A. acknowledges support by NAOJ ALMA Scientific Research Grant code 2019-13B, Grant-in-Aid for Scientific Research (S) 18H05222, and Grant-in-Aid for Transformative Research Areas (A) 20H05844 and 20H05847. G.C. is supported by the NAOJ ALMA Scientific Research Grant code 2019-13B. Y.Y. is supported by IGPEES, WINGS Program, The University of Tokyo.  K.Z. acknowledges the support of the Office of the Vice Chancellor for Research and Graduate Education at the University of Wisconsin – Madison with funding from the Wisconsin Alumni Research Foundation. S.M.A. and J.H. acknowledge funding support from the National Aeronautics and Space Administration under Grant No.\ 17-XRP17 2-0012 issued through the Exoplanets Research Program.
E.A.B. and A.D.B. acknowledges support from NSF AAG Grant \#1907653.  L.I.C. gratefully acknowledges support from the David and Lucille Packard Foundation and Johnson \& Johnson's WiSTEM2D Program. V.V.G. acknowledges support from FONDECYT Iniciaci\'on 11180904 and ANID project Basal AFB-170002. 
J.D.I. acknowledges support from the Science and Technology Facilities Council of the United Kingdom (STFC) under ST/T000287/1. C.J.L. acknowledges funding from the National Science Foundation Graduate Research Fellowship under Grant No.\ DGE1745303. R.L.G. acknowledges support from a CNES fellowship grant. F.M. acknowledges support from ANR of France under contract ANR-16-CE31-0013 (Planet-Forming-Disks) and ANR-15-IDEX-02 (through CDP ``Origins of Life"). H.N. acknowledges support by NAOJ ALMA Scientific Research Grant code 2018-10B and Grant-in-Aid for Scientific Research 18H05441. K.I.\"O. acknowledges support from the Simons Foundation (SCOL \#321183) and an NSF AAG Grant (\#1907653).  R.T. acknowledges support from the Smithsonian Institution as a Submillimeter Array (SMA) Fellow. T.T. is supported by JSPS KAKENHI Grant Numbers JP17K14244 and JP20K04017. C.W. acknowledges financial support from the University of Leeds, STFC and UKRI (grant numbers ST/R000549/1, ST/T000287/1, MR/T040726/1).
K.Z., K.R.S., J.H., J.B., J.B.B., and I.C. acknowledge the support of NASA through Hubble Fellowship grants HST-HF2-51401.001, HST-HF2-51419.001, HST-HF2-51460.001-A, HST-HF2-51427.001-A, HST-HF2-51429.001-A, and HST-HF2-51405.001-A awarded by the Space Telescope Science Institute, which is operated by the Association of Universities for Research in Astronomy, Inc., for NASA, under contract NAS5-26555.

%% To help institutions obtain information on the effectiveness of their telescopes the AAS Journals has created a group of keywords for telescope facilities. Following the acknowledgments section, use the following syntax and the \facility{} or \facilities{} macros to list the keywords of facilities used in the research for the paper.  Each keyword is check against the master list during copy editing. Individual instruments can be provided in parentheses, after the keyword, but they are not verified.
\vspace{5mm}
\facilities{ALMA}

%% Similar to \facility{}, there is the optional \software command to allow authors a place to specify which programs were used during the creation of the manuscript. Authors should list each code and include either a citation or url to the code inside ()s when available.
\software{Astropy \citep{astropy13,astropy18}, bettermoments \citep{Teague18_bettermoments}, CASA \citep{McMullin07}, emcee \citep{Foreman-Mackey13}, gofish \citep{Teague19_gofish}, matplotlib \citep{Hunter07}, NumPy \citep{vanderWalt11}, pythonradex (\url{https://github.com/gica3618/pythonradex}), SciPy \citep{Virtanen20}, SPECTCOL (VAMDC Consortium, \url{http://www.vamdc.org}), VISIBLE \citep{loomis18}.}

\appendix

\section{Hybrid zeroth moment maps with a 0\texorpdfstring{\farcs}{''}5 beam}\label{appendix:05taper}

\begin{figure}
\plotone{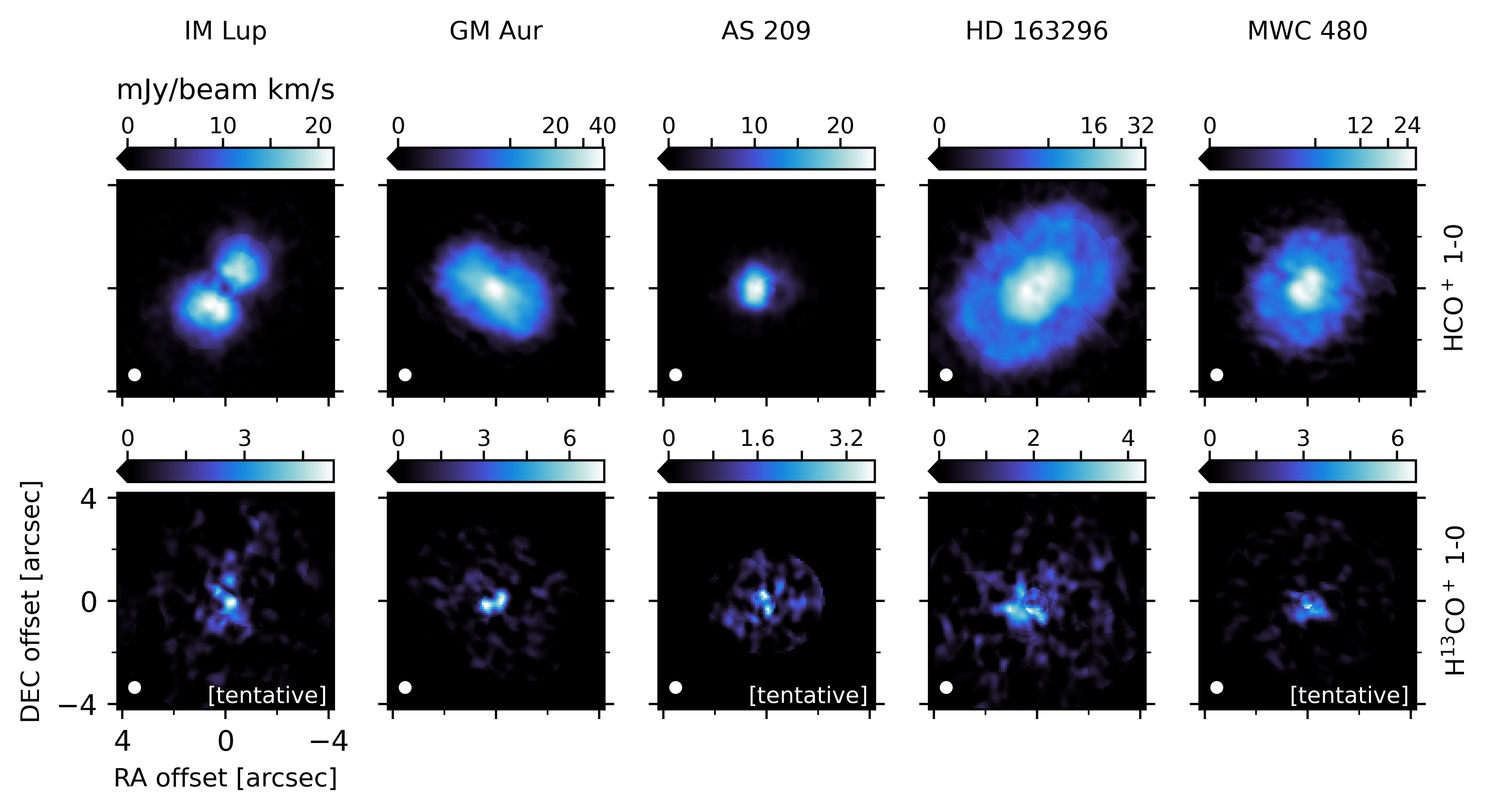}
\caption{Same as figure \ref{fig:HCO+_mom0_gallery}, but for images tapered to a circular 0.5\arcsec beam.}
\label{fig:mom0_05taper}
\end{figure}

While we used the data cubes tapered to a circular 0\farcs3 beam for the analysis, faint emission is better recognized in images with lower spatial resolution. 
Since Band 3 lines tend to be fainter than Band 6 lines, the MAPS collaboration also produced the data cube tapered to a circular 0\farcs5 beam for Band 3 lines \citep{oberg20,czekala20}. 
Figure \ref{fig:mom0_05taper} shows the hybrid zeroth moment maps of the $J=1-0$ lines of HCO$^+$ and H$^{13}$CO$^+$ tapered to a circular 0\farcs5 beam. All maps were produced by combining a Keplerian mask and a smoothed 0$\sigma$-clip mask \citep{law20_radial_profiles}.

\section{Matched filter analysis and disk-integrated spectra for \texorpdfstring{H$^{13}$CO$^+$}{H13CO+} \texorpdfstring{$J=1-0$}{J=1-0}}\label{appendix:matched_filter}

\begin{figure}
\plotone{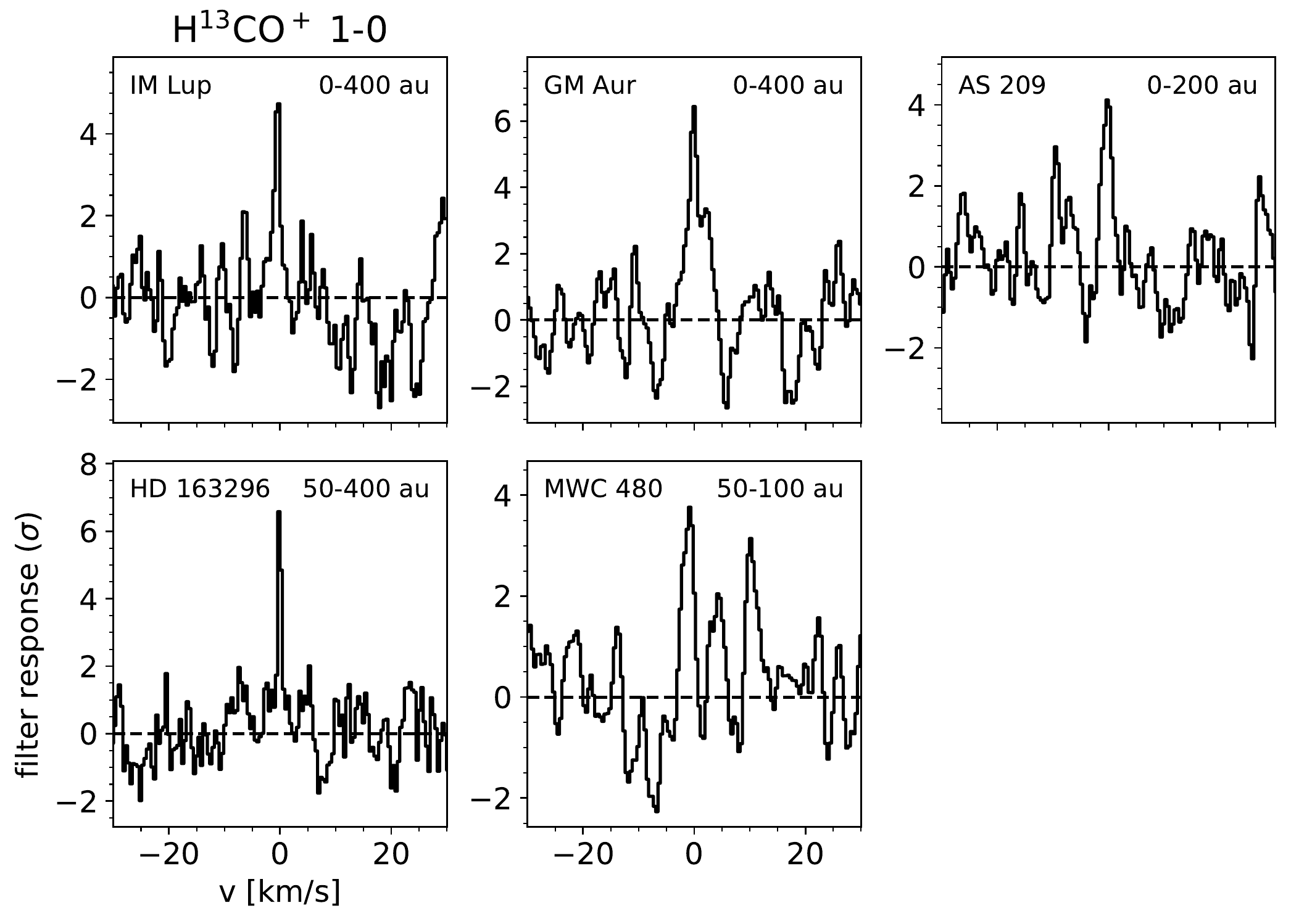}
\caption{Matched filter response for H$^{13}$CO$^+$ $J=1-0$. For each disk, we chose the Keplerian mask with the radius that maximized the SNR. The radial extent of that model is indicated in the upper right of each panel. The y-axis is in units of the noise in the filter response.  \label{fig:matched_filter}}
\end{figure}

The $J=1-0$ line of H$^{13}$CO$^+$ is not clearly seen in our zeroth moment maps (Figure \ref{fig:HCO+_mom0_gallery}), radial emission profiles (Figure \ref{fig:radial_profile}), or azimuthally averaged spectra
%(Figures \ref{fig:IM_Lup_az_spectra} to \ref{fig:MWC_480_az_spectra}). 
(Figure \ref{fig:IM_Lup_az_spectra}). 
To test whether the line is detected, we applied a matched filter analysis in the $uv$ plane \citep{loomis18}\citep[see also][]{czekala20}. 
We calculate the visibilities of a Keplerian disk model, which are then cross-correlated with the observed visibilities to produce a response for each channel of the data. The response spectrum is then divided by the rms of the response spectrum at signal-free regions. Since we do not know the radial distribution of the emission, 
we computed the filter response for a series of Keplerian disk models with varying radial extents to find the Keplerian model that maximizes the SNR at the systemic velocity. Figure \ref{fig:matched_filter} shows the matched filter responses with the maximum SNR, while the radial extent of the corresponding Keplerian disk models is shown in the figure and also listed in Table \ref{tab:disk_integrated_fluxes}. We note that the radial extent of the keplerian disk model roughly represents the radial extent of the emission.
H$^{13}$CO$^+$ $J=1-0$ is detected with a SNR above 6$\sigma$ for GM~Aur and HD~163296. The line is tentatively detected with a SNR above $3\sigma$ for IM~Lup, AS~209 and MWC~480.

\begin{figure}
\plotone{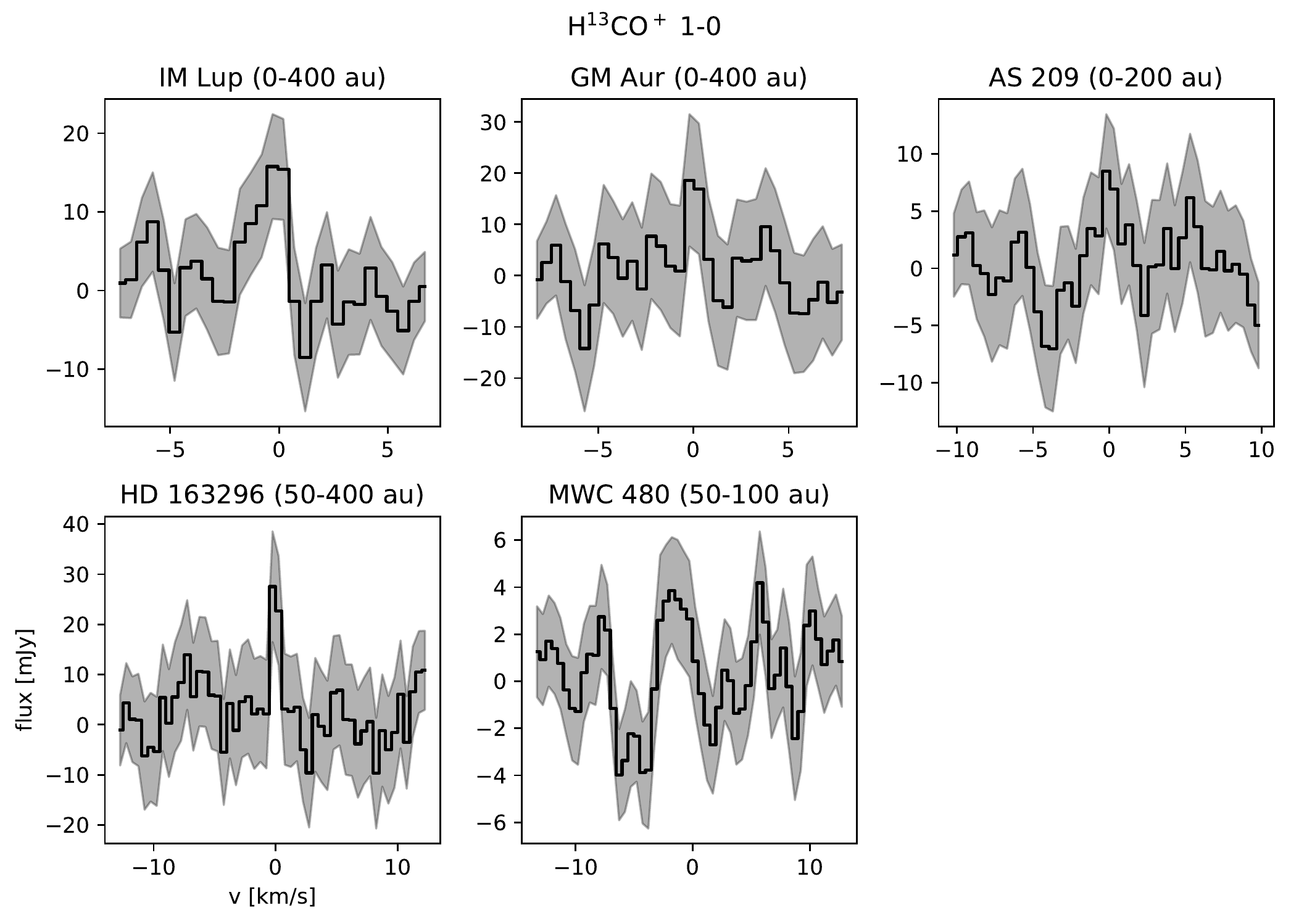}
\caption{Disk-integrated spectra (over the radial range indicated in the title of each panel) of H$^{13}$CO$^+$ $J=1-0$. The spectrum at each spatial location of the data cube was shifted to the systemic velocity prior to integration. The shaded region corresponds to the 1$\sigma$ error. \label{fig:disk_integrated_spectra_H13COp}}
\end{figure}

In Figure \ref{fig:disk_integrated_spectra_H13COp} we show the shifted and azimuthally averaged, disk-integrated H$^{13}$CO$^+$ $J=1-0$ spectrum of each disk. The radial extent of the region over which we averaged is the same as for the matched filter. The disk-integrated spectra also point towards a detection in GM~Aur and HD~163296, although at lower significance than the matched filter. In summary, we detect H$^{13}$CO$^+$ $J=1-0$ towards GM~Aur and IM~Lup, and tentatively detect it towards the other three disks.

\section{Azimuthally averaged spectra}\label{appendix:shifted_stacked_spectra}
%Figures \ref{fig:IM_Lup_az_spectra} to \ref{fig:MWC_480_az_spectra} 
%Figure \ref{fig:IM_Lup_az_spectra} 
Figure 16
show the azimuthally averaged spectra of \ce{HCO+} and H$^{13}$CO$^+$ $J=1-0$ together with model spectra from the MCMC fits.

\begin{figure}
\plotone{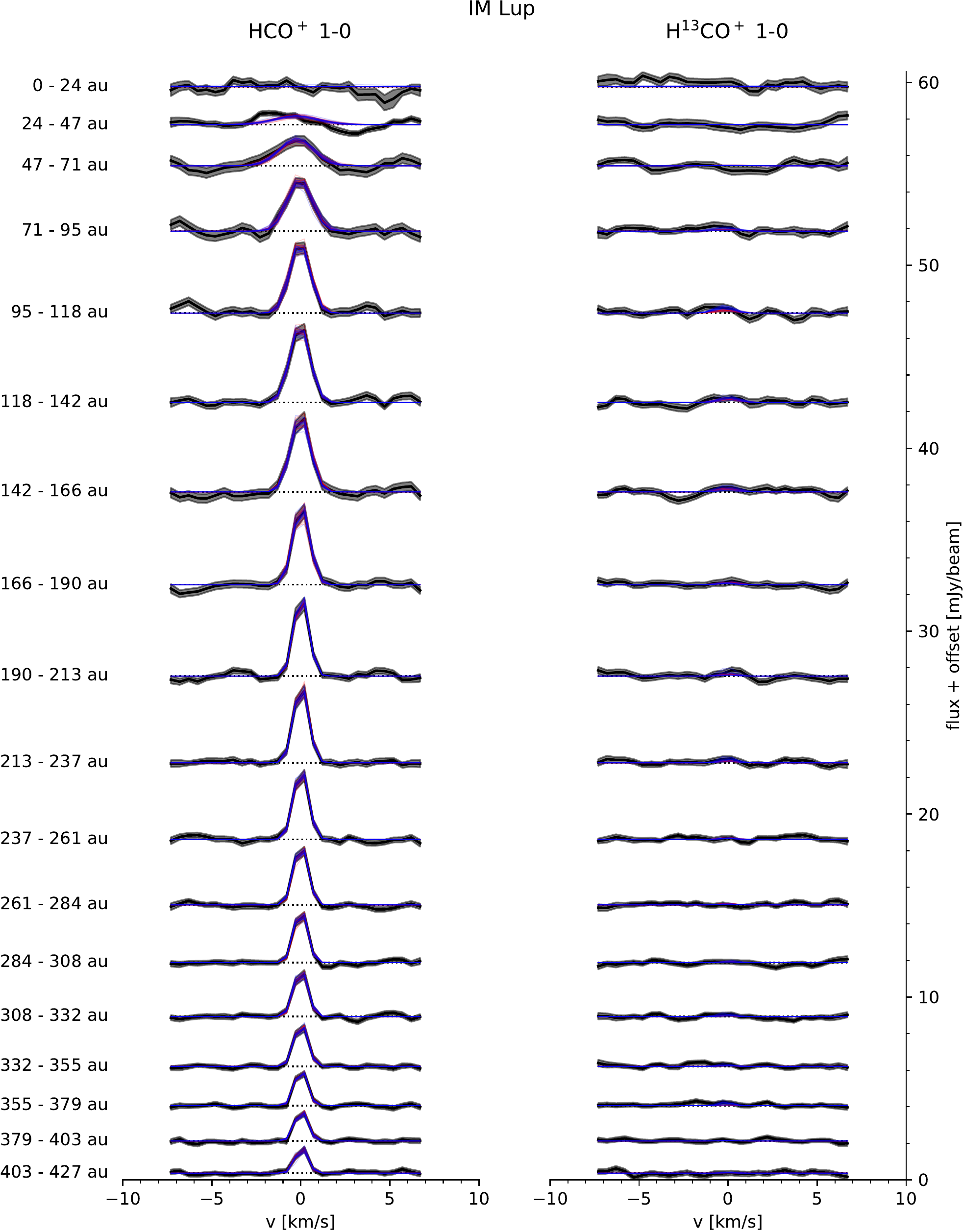}
\figurenum{16a}
\caption{The black curves show azimuthally averaged spectra for IM~Lup. Spectra are vertically offset for clarity. The shaded region marks the 1$\sigma$ error. The horizontal dotted line marks the zero flux level. For each spectrum, we show 20 randomly selected model spectra from the MCMC with the blue curves ($T_\mathrm{ex}$ as free parameter) and the red curves ($T_\mathrm{ex}$ fixed to 30\,K). The selection probability of each model was set proportional to its posterior probability. Most of the time, the model spectra overlap closely, so that only the blue curves are visible.
%The complete figure set (7 images) is available in the online journal. 
 \label{fig:IM_Lup_az_spectra}}
\end{figure}

\begin{figure}
\plotone{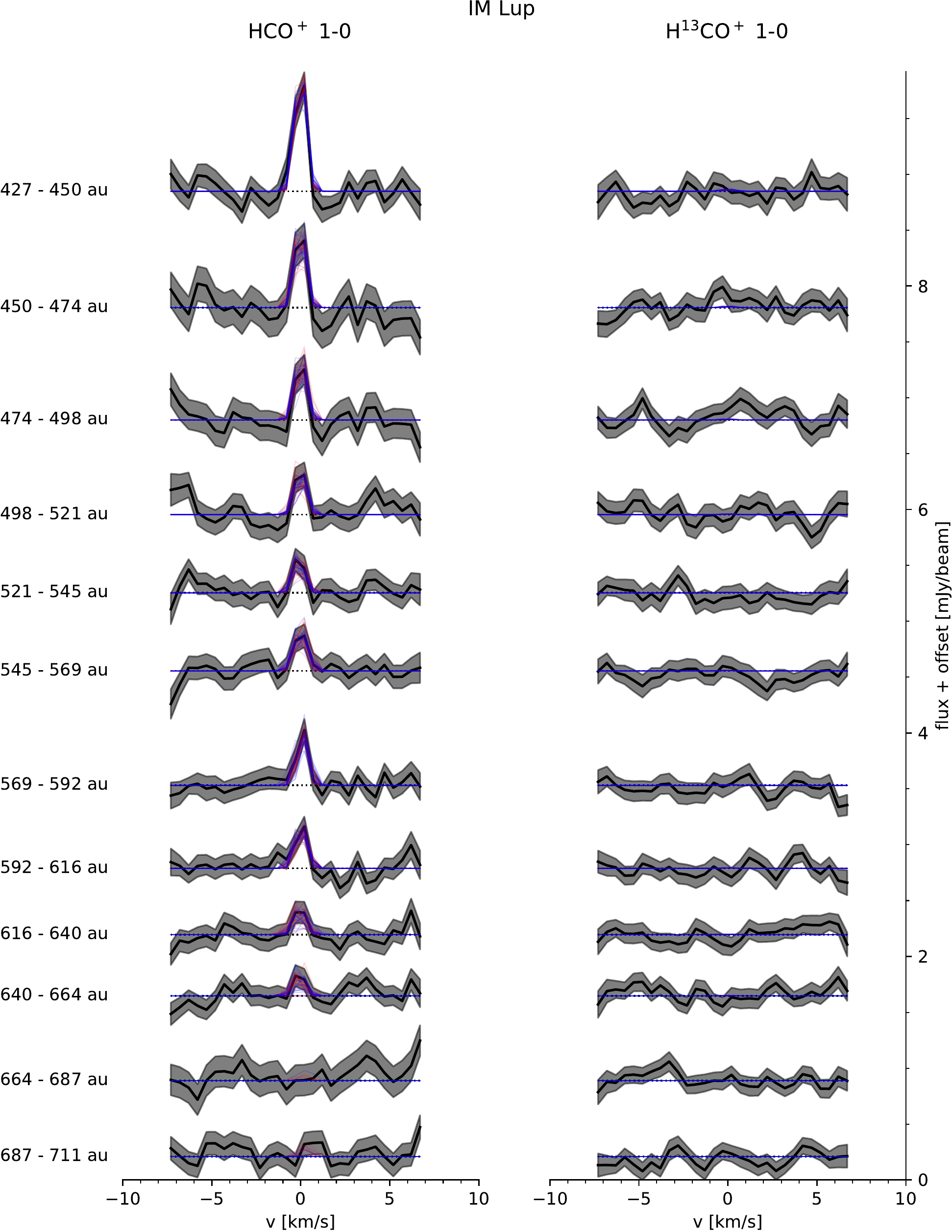}
\figurenum{16b}
\caption{Continuation of figure \ref{fig:IM_Lup_az_spectra}. \label{fig:IM_Lup_az_spectra_b}}
\end{figure}

\begin{figure}
\plotone{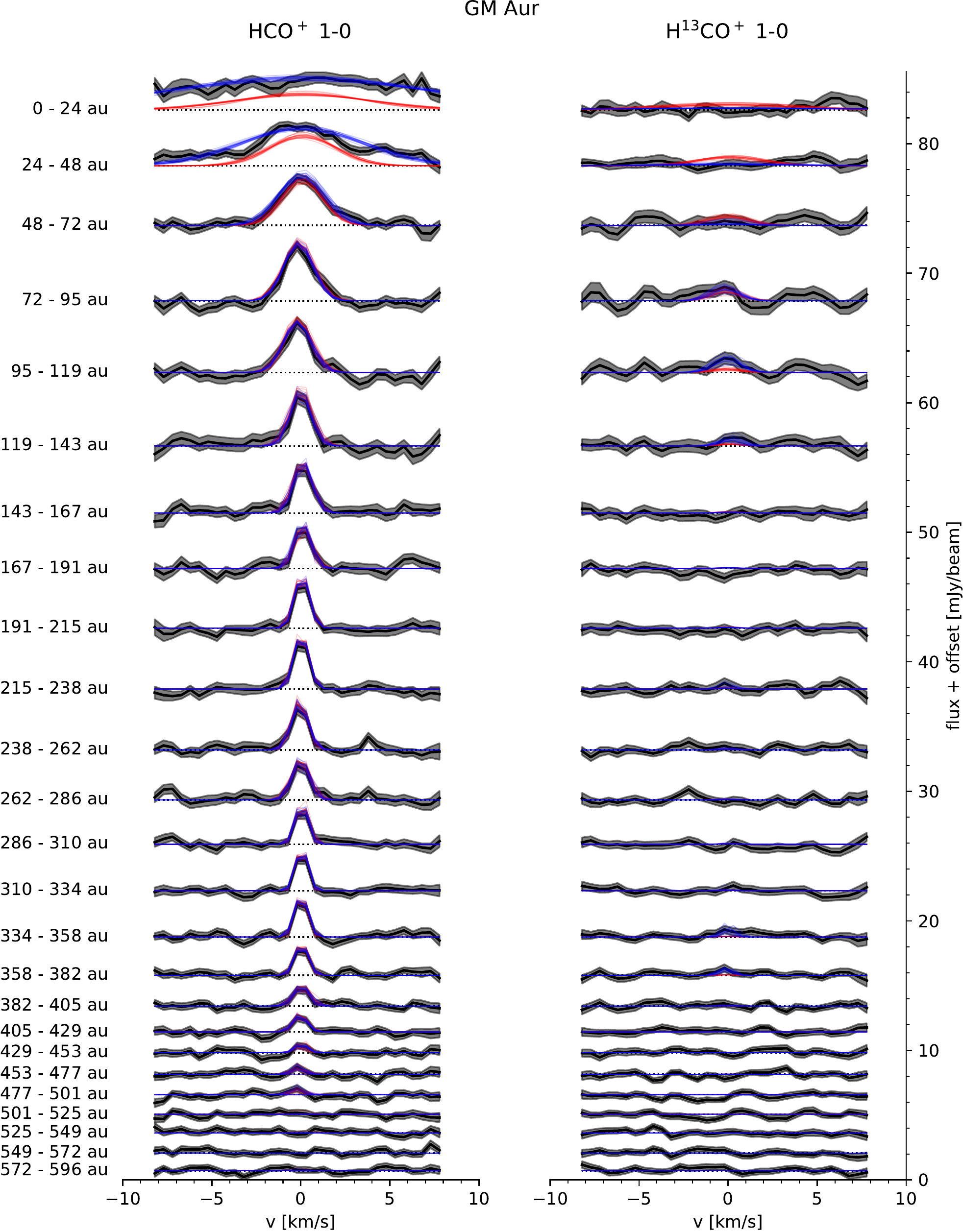}
\figurenum{16c}
\caption{Same as figure \ref{fig:IM_Lup_az_spectra}, but for GM~Aur. \label{fig:GM_Aur_az_spectra}}
\end{figure}

\begin{figure}
\plotone{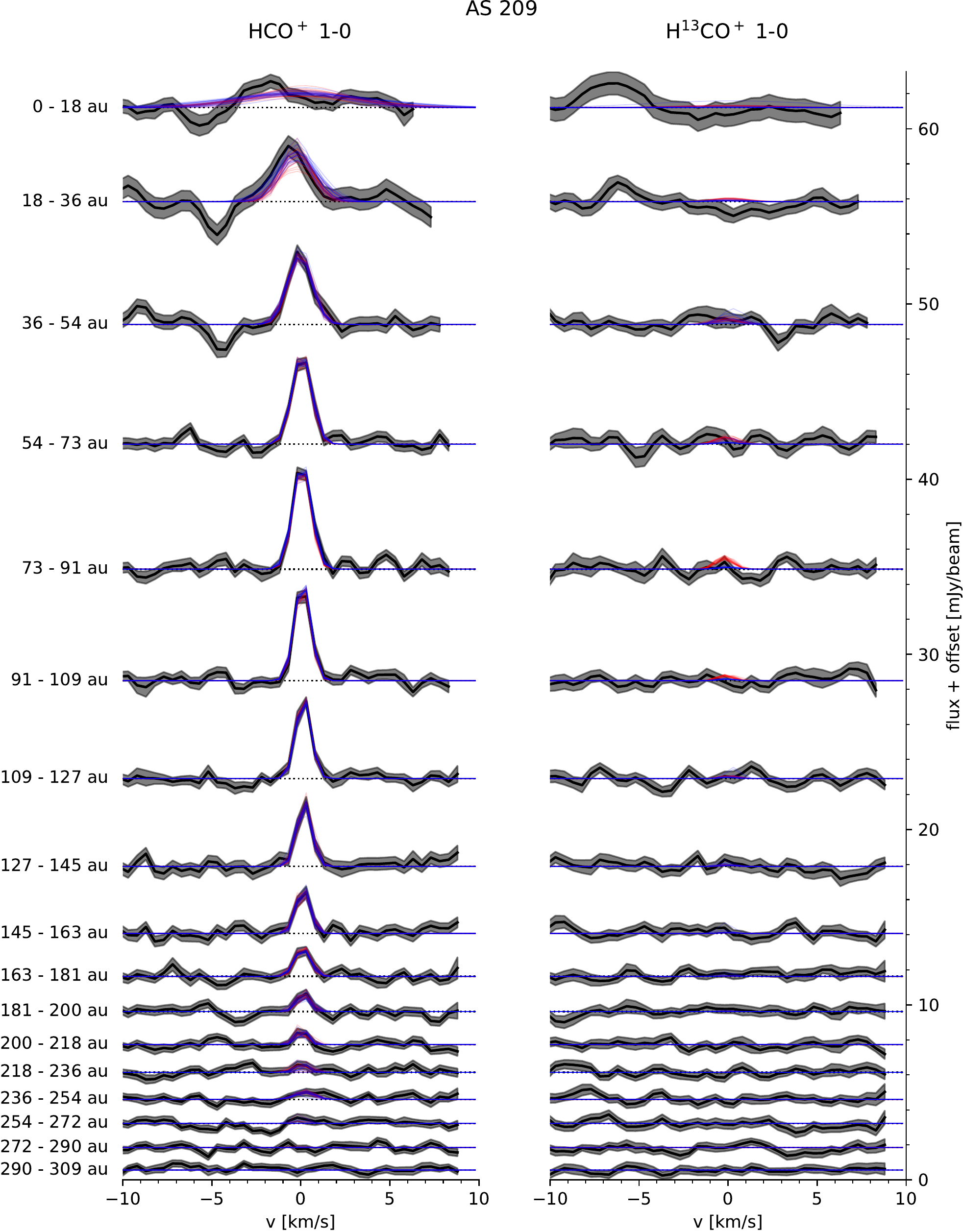}
\figurenum{16d}
\caption{Same as figure \ref{fig:IM_Lup_az_spectra}, but for AS~209. \label{fig:AS_209_az_spectra}}
\end{figure}

\begin{figure}
\plotone{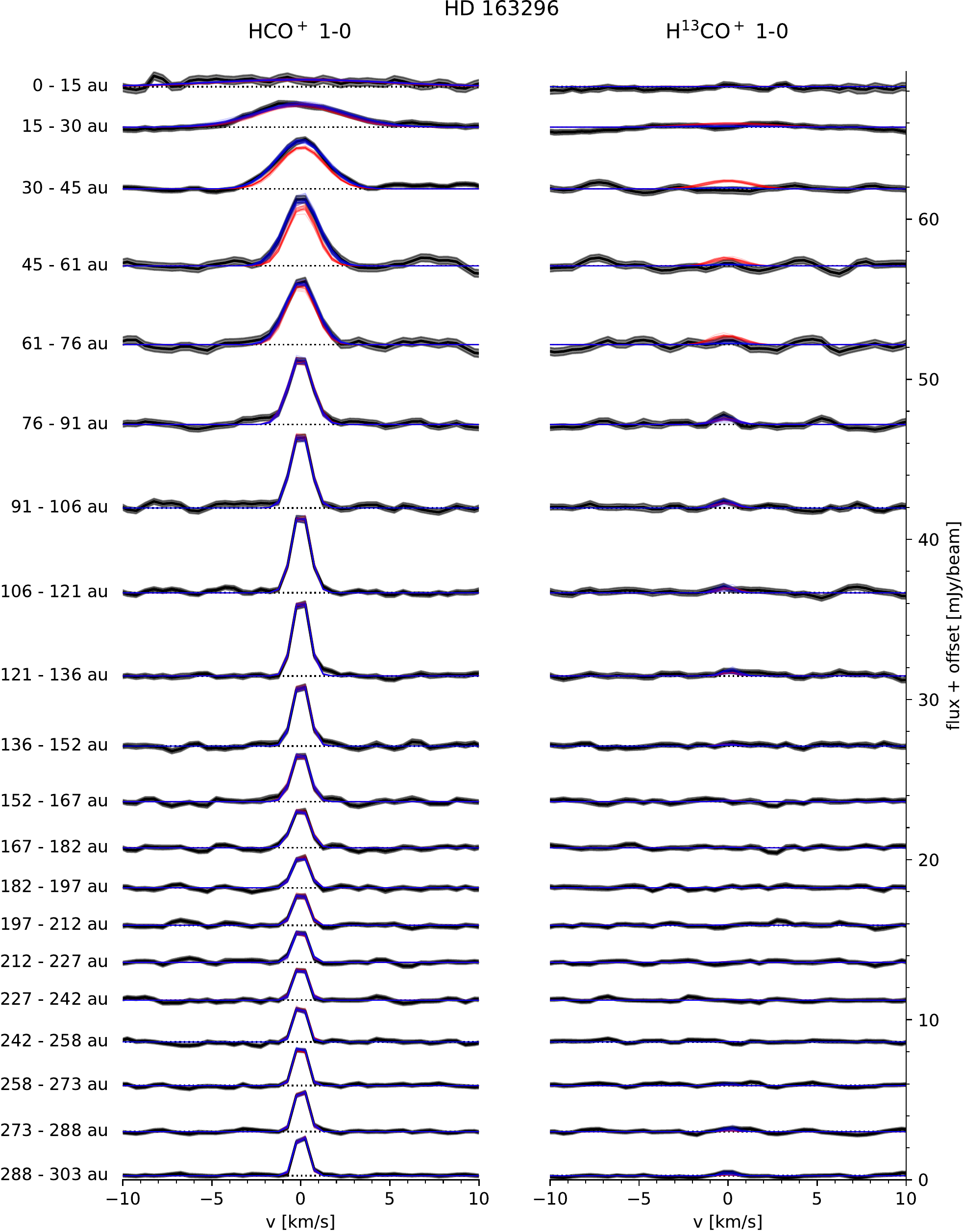}
\figurenum{16e}
\caption{Same as figure \ref{fig:IM_Lup_az_spectra}, but for HD~163296. \label{fig:HD_163296_az_spectra}}
\end{figure}

\begin{figure}
\plotone{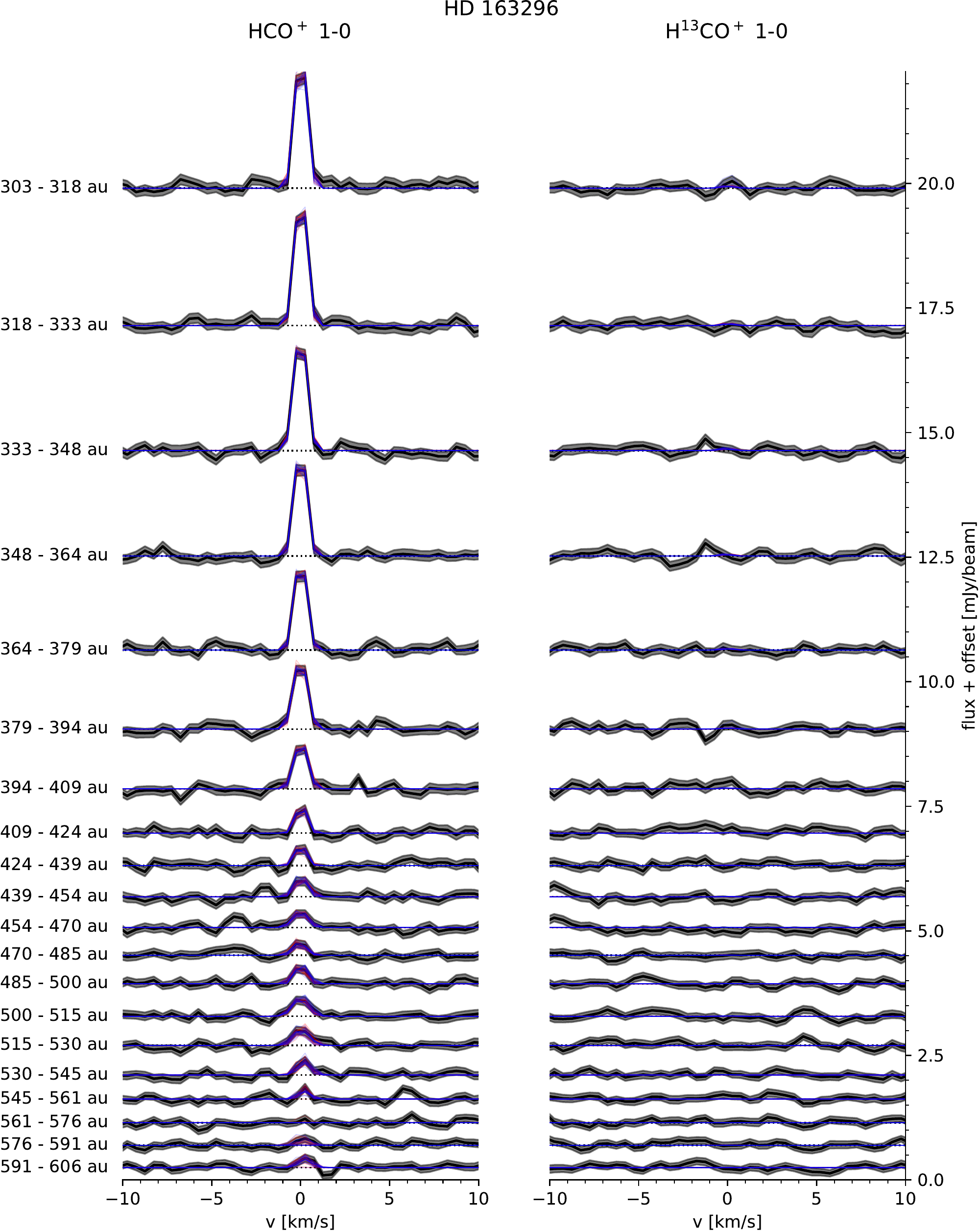}
\figurenum{16f}
\caption{Continuation of figure \ref{fig:HD_163296_az_spectra}. \label{fig:HD_163296_az_spectra_b}}
\end{figure}

\begin{figure}
\plotone{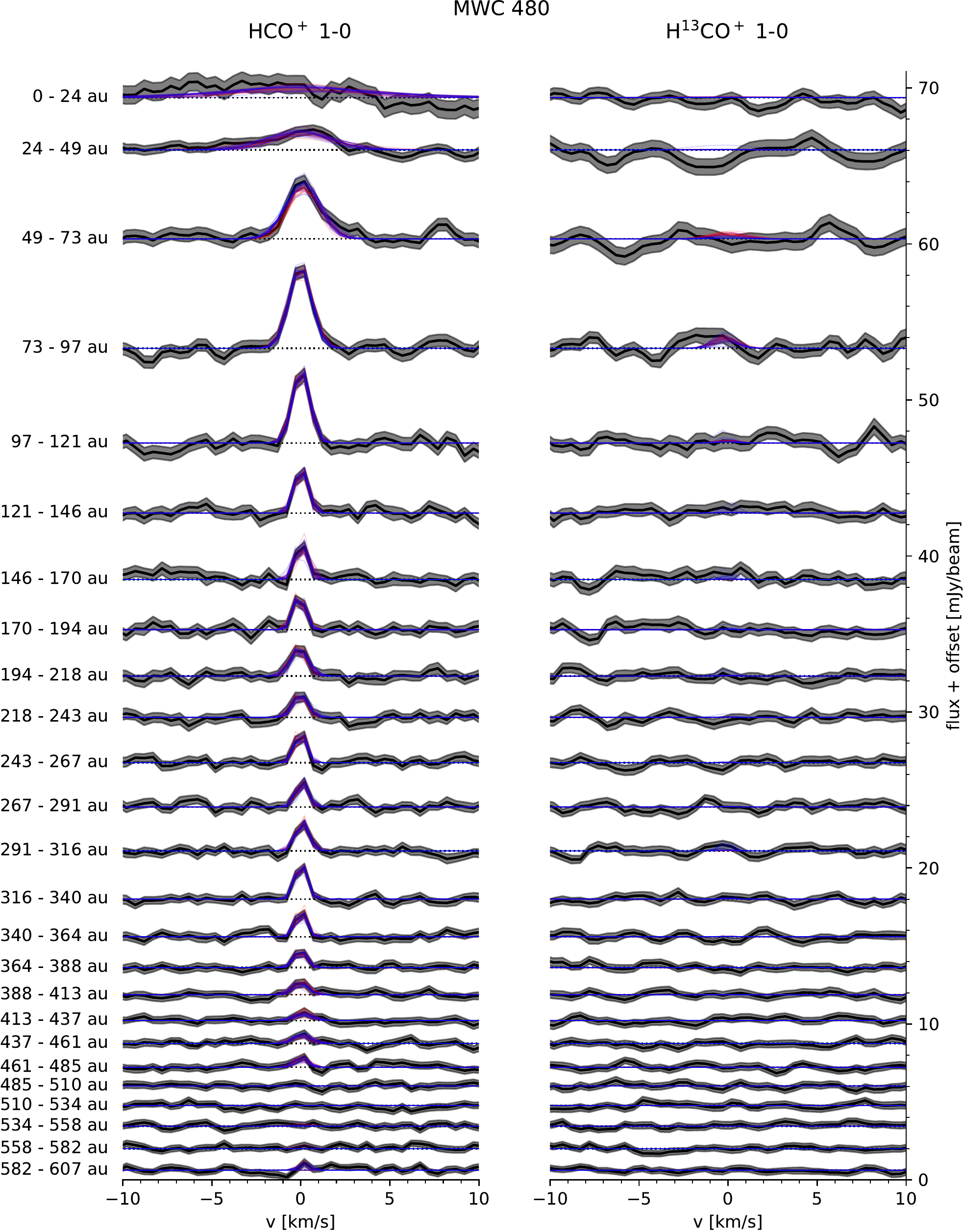}
\figurenum{16g}
\caption{Same as figure \ref{fig:IM_Lup_az_spectra}, but for MWC~480. \label{fig:MWC_480_az_spectra}}
\end{figure}

\section{Dependence of HCO$^+$ column density on assumed excitation temperature}\label{appendix:HCOp_column_density_T_dependence}

\begin{figure}
\plotone{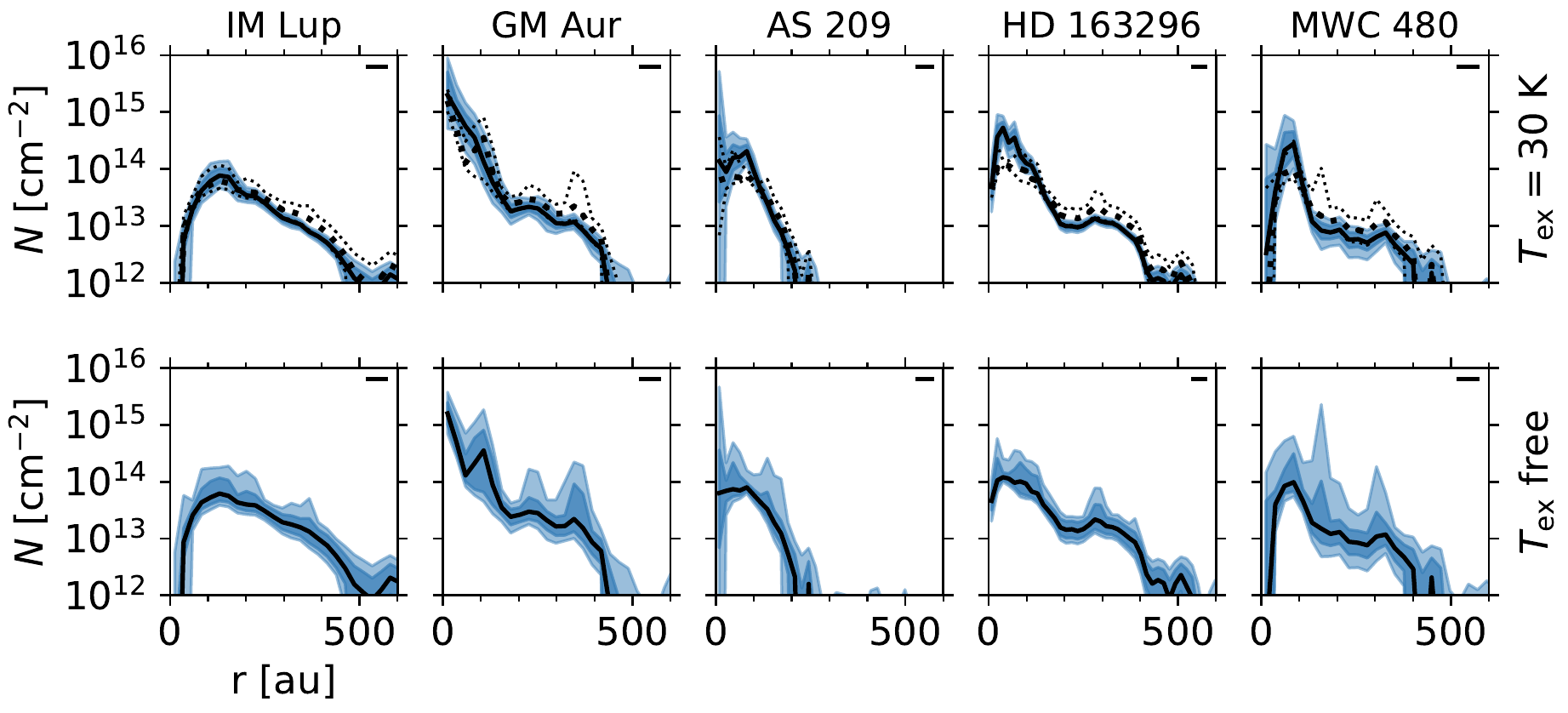}
\figurenum{17}
\caption{The column density of HCO$^+$ assuming an excitation temperature of 30\,K (top row) and for our fiducial fits where $T_\mathrm{ex}$ is a free parameter (bottom row). The black solid lines mark the median, while the shading encompasses the regions from the 16th to 84th and the 2.3th to 97.7th percentile, respectively. To ease comparison, in the top panel, the thick and thin black dotted lines mark the median and the 16th and 84th percentile of the fits where $T_\mathrm{ex}$ is a free parameter. \label{fig:N_HCO+_Tex30_comparison}}
\end{figure}

\begin{figure}
\plotone{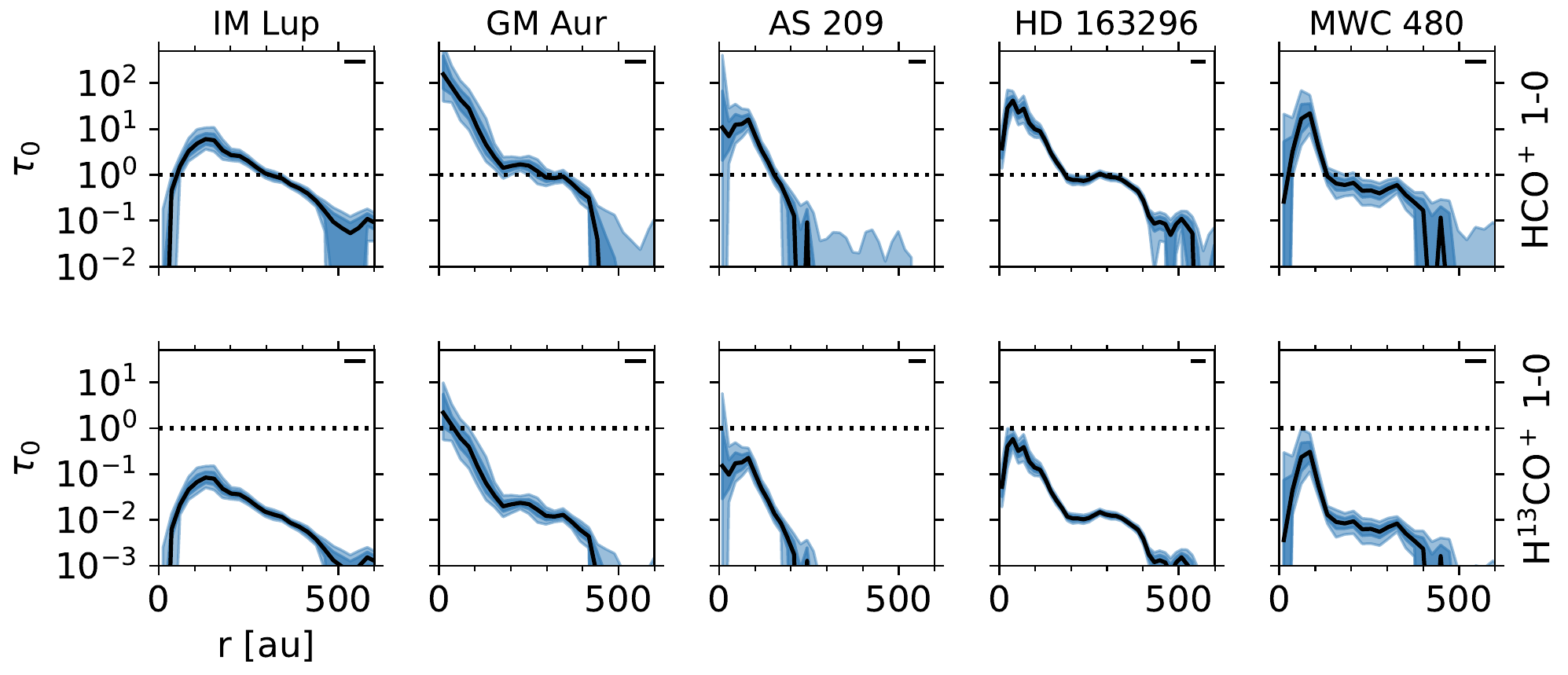}
\figurenum{18}
\caption{Optical depth profiles of HCO$^+$ ($J=1-0$) and H$^{13}$CO$^{+}$ ($J=1-0$) derived from fitting azimuthally averaged spectra assuming $T_\mathrm{ex}=30$\,K. The black lines mark the median, while the shading encompasses the regions from the 16th to 84th and the 2.3th to 97.7th percentile, respectively.\label{fig:HCOP_tau_Tex30}}
\end{figure}

In our fiducial fits, the excitation temperature is a free parameter. In order to explore the dependence of our results on the excitation temperature, we performed additional fits where the excitation temperature is fixed to 30\,K. Figure \ref{fig:N_HCO+_Tex30_comparison} displays a comparison of the column density profiles derived from the different fits. The comparison is discussed in Section \ref{section:column}. Figure \ref{fig:HCOP_tau_Tex30} shows the optical depth profiles derived assuming $T_\mathrm{ex}=30$\,K.

\section{Template disk chemistry models}\label{appendix:model}

\begin{figure}
\plotone{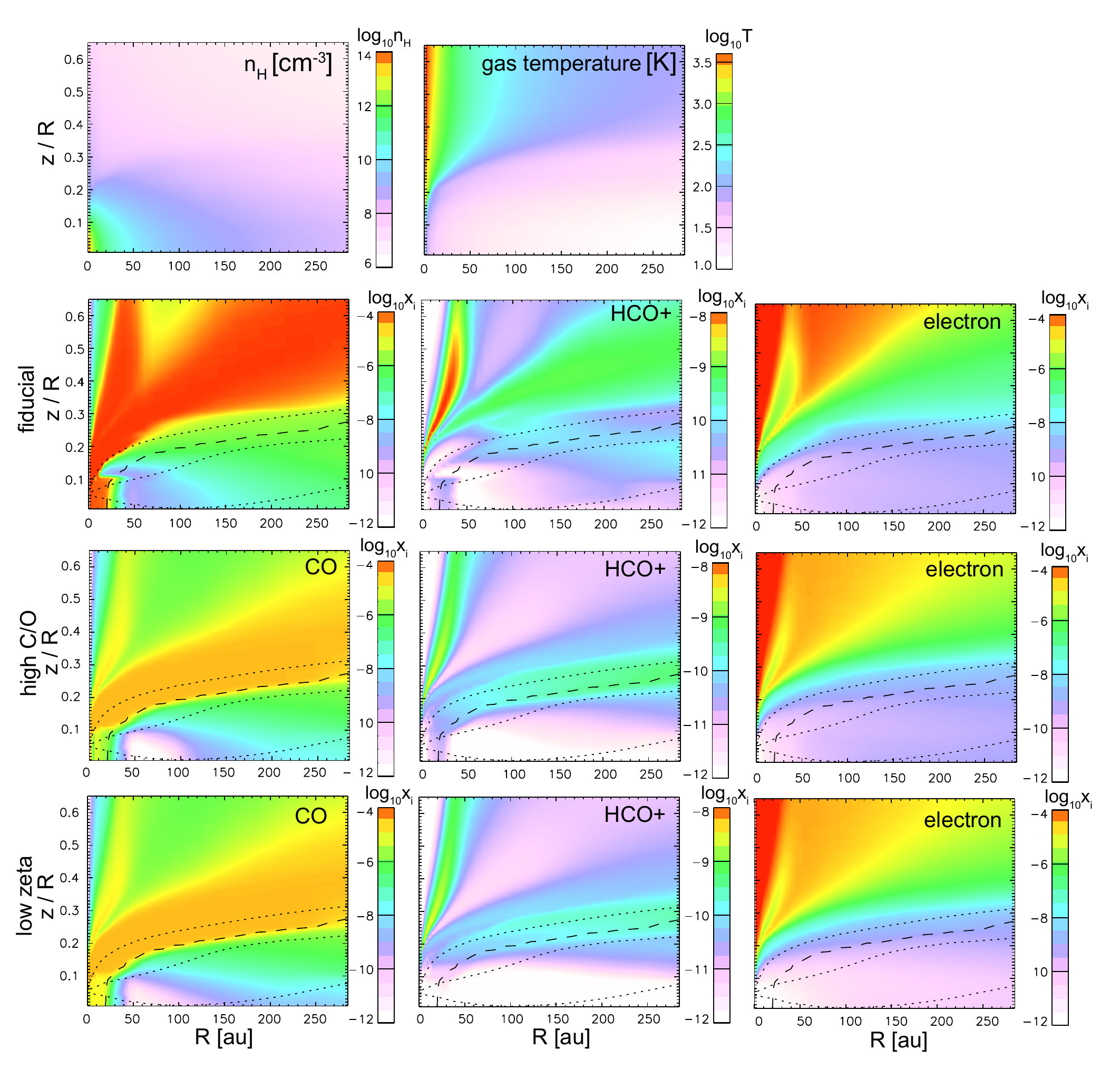}
\figurenum{19}
\caption{Number density of hydrogen nuclei ($n_{\rm H}$) and gas temperature of the template disk model are shown in the top row. Furthermore, we show the abundances of CO (left column), HCO$^+$ (middle column), and electrons (right column) relative to hydrogen nuclei in the fiducial model (second row), CO depleted model (third row), and the model without CR ionization (low $\zeta$ model, bottom row).
\label{fig:model_rel_abun_appendix}}
\end{figure}

To top panels in Figure \ref{fig:model_rel_abun_appendix} show the 2D ($R$, $Z$) distributions of the number of hydrogen nuclei and gas temperature in the template disk model. The panels in the second row show the abundances of gaseous CO, HCO$^+$, and electron relative to hydrogen nuclei in the fuducial model. The panels in the third and fourth row are for the high C/O model and the low $\zeta$ model.

\begin{figure}
\plotone{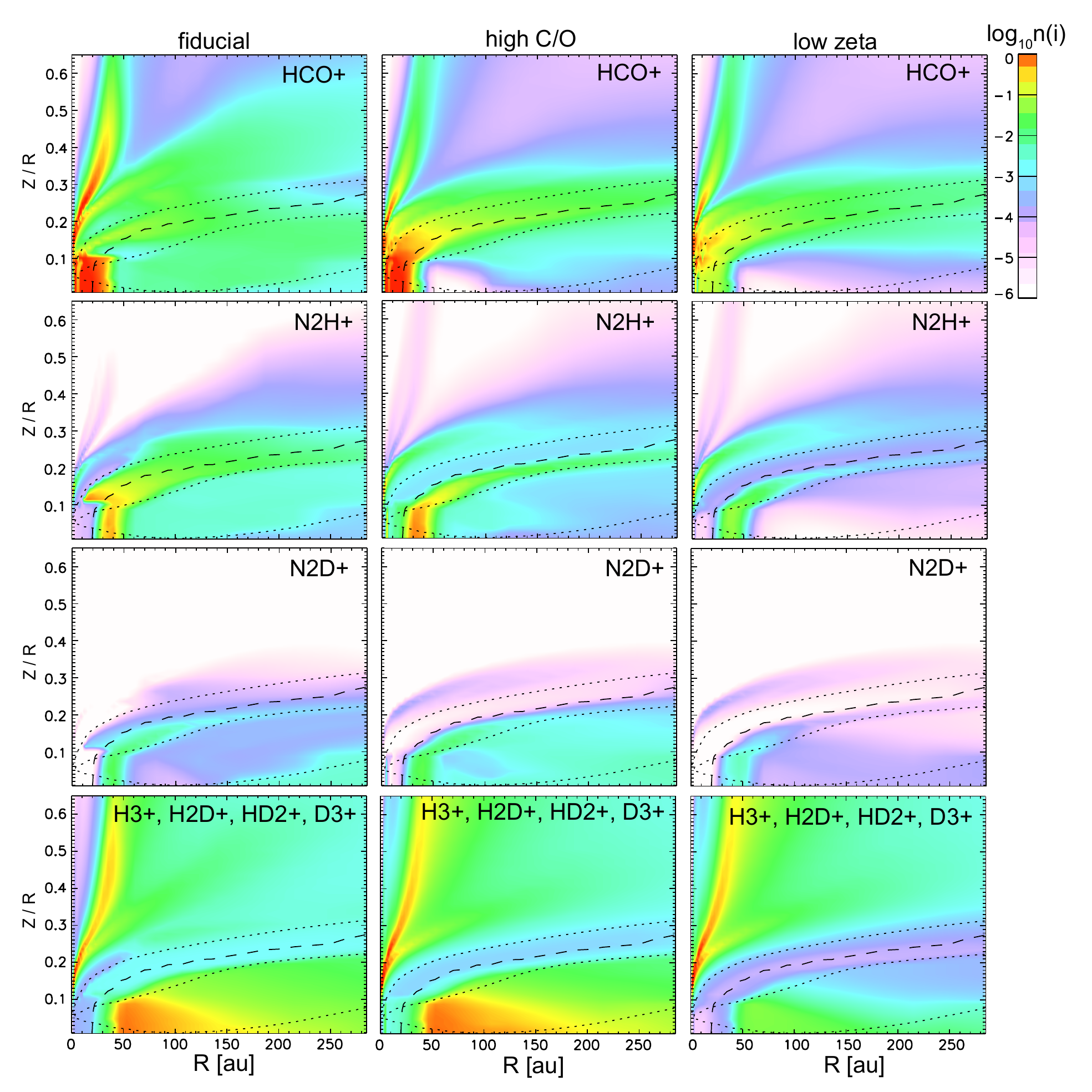}
\figurenum{20}
\caption{
Absolute abundance (i.e.\ number density in units of cm$^{-3}$) of major molecular ions in the fiducial disk model (left column),
high C/O model (middle column), and low $\zeta$ model (right column).
\label{fig:model_abs_ion_appendix}}
\end{figure}

Figure \ref{fig:model_abs_ion_appendix} shows the absolute abundances (i.e. number density) of HCO$^+$ (top row), N$_2$H$^+$ (the second row), N$_2$D$^+$ (the third row), H$_3^+$ and its deuterated isotopomers (the fourth row) in the fiducial model (left), the high C/O model (middle), and the low $\zeta$ model (right).

\bibliography{bibliography}

\end{document}